\documentclass[journal=jacsat,manuscript=article]{achemso}

\usepackage[version=3]{mhchem}
\usepackage{amsmath}
\usepackage{amssymb}
\usepackage{graphicx}
\usepackage{booktabs}
\usepackage{multirow}
\usepackage{listings}
\usepackage{xcolor}
\usepackage{subcaption}
\usepackage{hyperref}

\newcommand{\ptychi}{\textsc{Ptychi-Evolve}}
\newcommand{\bfr}{\mathbf{r}}
\newcommand{\bfq}{\mathbf{q}}

\author{Xiangyu Yin}
\email{xyin@anl.gov}
\affiliation[Argonne National Laboratory]{Advanced Photon Source, Argonne National Laboratory, Lemont, IL, USA}

\author{Ming Du}
\affiliation[Argonne National Laboratory]{Advanced Photon Source, Argonne National Laboratory, Lemont, IL, USA}

\author{Junjing Deng}
\affiliation[Argonne National Laboratory]{Advanced Photon Source, Argonne National Laboratory, Lemont, IL, USA}

\author{Zhi Yang}
\affiliation[Rice University]{Department of Materials Science and NanoEngineering, Rice University, Houston, TX, USA}

\author{Yimo Han}
\affiliation[Rice University]{Department of Materials Science and NanoEngineering, Rice University, Houston, TX, USA}

\author{Yi Jiang}
\affiliation[Argonne National Laboratory]{Advanced Photon Source, Argonne National Laboratory, Lemont, IL, USA}

\title{Autonomous Algorithm Discovery for Ptychography via Evolutionary LLM Reasoning}

\keywords{ptychography, large language models, algorithm discovery, regularization, evolutionary computation, computational imaging}

\makeatletter
\let\acs@delayed@maketitle\maketitle
\renewcommand*{\maketitle}{} 
\makeatother

\begin{document}

\pagenumbering{gobble}
\thispagestyle{empty}
\textbf{GOVERNMENT LICENSE}

The submitted manuscript has been created by UChicago Argonne, LLC, Operator of Argonne National Laboratory (“Argonne”). Argonne, a U.S. Department of Energy Office of Science laboratory, is operated under Contract No. DE-AC02-06CH11357. The U.S. Government retains for itself, and others acting on its behalf, a paid-up nonexclusive, irrevocable worldwide license in said article to reproduce, prepare derivative works, distribute copies to the public, and perform publicly and display publicly, by or on behalf of the Government. The Department of Energy will provide public access to these results of federally sponsored research in accordance with the DOE Public Access Plan. \href{http://energy.gov/downloads/doe-public-access-plan}{http://energy.gov/downloads/doe-public-access-plan}
\clearpage
\pagenumbering{arabic}

\makeatletter
\let\maketitle\acs@delayed@maketitle
\makeatother
\maketitle

\begin{abstract}
Ptychography is a computational imaging technique widely used for high-resolution materials characterization, but high-quality reconstructions often require the use of regularization functions that largely remain manually designed. We introduce \ptychi{}, an autonomous framework that uses large language models (LLMs) to discover and evolve novel regularization algorithms. The framework combines LLM-driven code generation with evolutionary mechanisms, including semantically-guided crossover and mutation. Experiments on three challenging datasets (X-ray integrated circuits, low-dose electron microscopy of apoferritin, and multislice imaging with crosstalk artifacts) demonstrate that discovered regularizers outperform conventional reconstructions, achieving up to +0.26 SSIM and +8.3~dB PSNR improvements. Besides, \ptychi{} records algorithm lineage and evolution metadata, enabling interpretable and reproducible analysis of discovered regularizers.
\end{abstract}

\section{Introduction}

Autonomous scientific discovery, in which artificial intelligence (AI) systems independently formulate hypotheses, design experiments, and interpret results, has long captivated researchers. Recent advances in large language models (LLMs) have brought this vision closer to reality, with systems demonstrating fully automated computational research pipelines\cite{lu2024aiscientist} and LLM-guided evolutionary search discovering novel algorithms in mathematics and computer science.\cite{romeraparedes2024funsearch,novikov2025alphaevolve} A central insight from those work is that searching through program space, by evolving code that describes \emph{how} to construct solutions, yields interpretable discoveries that can be verified, generalized, and deployed in practice.

Ptychography presents a good domain for such autonomous algorithm discovery. This computational imaging technique reconstructs high-resolution images from overlapping diffraction patterns, achieving resolution beyond conventional optical limits.\cite{thibault2008highres, pfeiffer2018xray,chen2021electron} The technique has become indispensable for nanoscale materials characterization, with applications spanning battery electrodes, catalytic nanoparticles, semiconductor devices, and biological specimens. However, ptychographic reconstruction is fundamentally an ill-posed and mostly underdetermined inverse problem whose quality depends critically on regularization, i.e., mathematical constraints encoding prior knowledge and limitations about the expected solution.\cite{marchesini2007unified} Despite decades of research producing diverse approaches from classical total variation (TV) to modern deep learning priors, the design of effective regularization strategies remains largely manual, requiring practitioners to select among competing methods and tune hyperparameters for each new imaging challenge. While recent work has automated hyperparameter selection vis Bayesian optimization \cite{cao2022automatic} and LLM-assisted automated workflows\cite{yin2024pearfull}, these approaches operate within fixed algorithmic structures rather than discovering new regularization strategies.

In this work, we propose \ptychi{}, an autonomous framework that uses LLM reasoning to discover and evolve regularization algorithms for the ptychographic reconstruction engine implemented in the Pty-Chi package.\cite{du2025ptychi} Unlike prior work that operates within fixed structures, \ptychi{} searches the space of algorithms themselves, generating executable Python functions that implement novel regularization strategies. By combining LLM-driven code generation with evolutionary refinement mechanisms including semantically-guided crossover and mutation, \ptychi{} discovers regularizers that substantially outperform conventional approaches while preserving interpretable algorithm lineage and evolution metadata.

The contributions of this work are fourfold. First, we present the first framework for autonomous discovery of regularization algorithms for computational imaging, extending beyond parameter optimization to structural algorithm search. Second, we introduce semantically-guided evolutionary operations where crossover and mutation are informed by the LLM's understanding of algorithmic function rather than random recombination. Third, we develop a flexible evaluation pipeline supporting ground-truth metrics, human expert feedback, and vision-language model assessment, enabling deployment across diverse experimental contexts. Fourth, we demonstrate substantial improvements across three challenging case studies, with analysis of evolution histories revealing interpretable scientific insights about effective regularization strategies.

\section{Related Work}

\subsection{Large Language Models for Scientific Discovery}

The application of large language models to scientific discovery has emerged as a rapidly evolving research frontier. For example, the AI Scientist system from Sakana AI represents one of the first efforts toward fully automated research, demonstrating end-to-end pipelines that generate novel research ideas, write necessary code, execute computational experiments, and produce complete scientific manuscripts.\cite{lu2024aiscientist} 

Within the specific domain of algorithmic discovery, FunSearch established the paradigm of searching through program space using LLM-guided evolution.\cite{romeraparedes2024funsearch} The key insight is that programs describing how to construct solutions are more amenable to evolutionary improvement than raw solutions, and the resulting discoveries are inherently interpretable. AlphaEvolve extends this approach by employing an ensemble of Gemini 2.0 models and scaling to programs hundreds of lines long.\cite{novikov2025alphaevolve} The system has achieved remarkable results including improvements to fundamental algorithms, optimization of Google's data center scheduling, and acceleration of TPU arithmetic circuits. The Evolution of Heuristics (EoH) framework introduced the concept of evolving both natural language ``thoughts'' representing algorithmic ideas and executable code implementing those ideas.\cite{liu2024eoh} This dual representation enables the LLM to reason about algorithmic concepts at a high level while still producing concrete, evaluable implementations. Similarly, the LLaMEA framework similarly uses LLMs for automated generation and refinement of metaheuristic algorithms, achieving performance exceeding state-of-the-art methods like CMA-ES and Differential Evolution.\cite{vanstein2024llamea}

Beyond algorithmic discovery, LLMs have found numerous applications in natural sciences like materials science and chemistry. We refer readers to a comprehensive survey documenting 34 examples spanning automation, assistants, agents, and accelerated discovery.\cite{zimmermann2025llm} In microscopy, the AILA framework enables LLM agents to automate microscopy experiments in user facilities\cite{prince2024opportunities}, and the PEAR framework \cite{yin2024pearfull} enables automated microscopy data analysis workflows. These efforts collectively establish that LLM agents can meaningfully engage with scientific context and analysis, though significant challenges remain.

\subsection{Ptychography reconstruction algorithms and artifacts}

A coherent diffraction imaging technique like ptychography collects diffraction patterns, and relies on an inverting process known as phase retrieval to recover human-interpretable images of the measured specimen. Examples of phase retrieval algorithms include the ptychographic iterative engine (PIE) family \cite{Rodenburg2004-nw, Maiden2009-md, Maiden2017-um}, least-square maximum likelihood (LSQML) \cite{Odstrcil2018-ns}, and difference map (DM) \cite{Elser2003-ov}. Software packages implementing these algorithms are available with various programming frameworks and offer a wide collection of features to enhance the reconstruction quality \cite{Wakonig2020-ap, Nashed2014-eh, Yue2021-ag, Enders2016-vs,Favre-Nicolin2020-cj}. Pty-Chi \cite{du2025ptychi} is a recent ptychographic data processing library leveraging PyTorch \cite{Paszke2019-xm} for numerical computing, automatic differentiation, GPU acceleration, and multi-processing. Pty-Chi also features a modular design that makes it possible to modify and improve certain features via drop-in replacement, facilitating the dynamic evolution of the reconstruction workflow. 

Ptychography is very powerful, but artifacts often arise from model mismatch and experimental imperfections.\cite{guizarsicairos2021ptychography} For example, periodic artifacts from raster-scan geometry, also known as grid pathology, are especially prominent when scan step sizes are large relative to the illumination function.\cite{hoidn2023gridartifacts} These artifacts originate from periodically repeating scan grids, aliasing in undersampled overlap regions, and systematic probe variations. While denser sampling suppresses such artifacts, this increases acquisition time and radiation dose, motivating algorithmic solutions. For thick specimens, multislice ptychography reconstructs depth-resolved transmission functions but suffers from crosstalk where features leak between slices.\cite{maiden2012multislice,suzuki2014multislice,oleary2024buried,du2021ddip} For dose-sensitive biological or rare samples, low-dose acquisition is often necessary to mitigate radiation damage. Under low-dose conditions, shot-noise–limited diffraction patterns and the reduced overlap common in sparse scanning weaken the reconstruction constraints, leading to grainier, lower-resolution images, while detector readout noise, background offsets, and unmodeled static intensity can further bias the reconstruction and introduce additional artifacts.\cite{seifert2023maximum, leidl2024influence, wu2024dose}

\subsection{Regularization in Ptychography}

Regularization is fundamental to solving ill-posed inverse problems in imaging. For example, the total variation (TV) functional, introduced by Rudin, Osher, and Fatemi, promotes edge-preserving solutions by penalizing the $L_1$ norm of image gradients.\cite{rudin1992tv} This approach along with its extensions including total generalized variation (TGV),\cite{bredies2010tgv}, have become ubiquitous in medical imaging, microscopy, and remote sensing. Anisotropic regularization approaches adapt penalty strength based on local image structure. Structure tensor analysis computes local orientation fields that guide preferential smoothing along edges rather than across them.\cite{weickert1998anisotropic} For images with regular structures, such as the metal lines in integrated circuits, orientation-aware regularization can improve reconstruction quality by respecting the underlying geometry.

The deep learning revolution has transformed regularization for inverse problems. The Deep Image Prior demonstrated that untrained convolutional networks provide implicit regularization through architectural inductive bias.\cite{ulyanov2018dip} This result showed that the structure of deep networks inherently favors natural images over noise, allowing reconstruction without any training data. Plug-and-play methods integrate learned denoisers into classical optimization frameworks, with theoretical foundations established through connections to proximal operators and ADMM.\cite{venkatakrishnan2013pnp,zhang2021pnp} The approach has achieved state-of-the-art results across deblurring, super-resolution, and computed tomography reconstruction.\cite{kamilov2023pnp}

For ptychography specifically, regularization choices significantly impact reconstruction quality. Classical approaches apply TV or sparsity constraints to the recovered object.\cite{maiden2009improved} Design of those regularization strategies is largely manual, requiring expert knowledge to select and tune approaches for specific imaging conditions. Recent work has explored learned approaches including deep image prior and deep generative priors.\cite{du2021ddip, aslan2021joint, barutcu2022compressive, cam2025diffusion} Du et al.\cite{du2026fidelity} developed a plug-and-play framework integrating physics-based reconstruction with text-guided diffusion models, enabling users to specify artifacts for removal through natural language prompts while maintaining data fidelity via ADMM. Those learned approaches aren't manual, but it's not interpretable and the performance is largely uncertain.

\section{Method: The Ptychi-Evolve Framework}

\begin{figure}[htbp]
\centering
\includegraphics[width=\columnwidth]{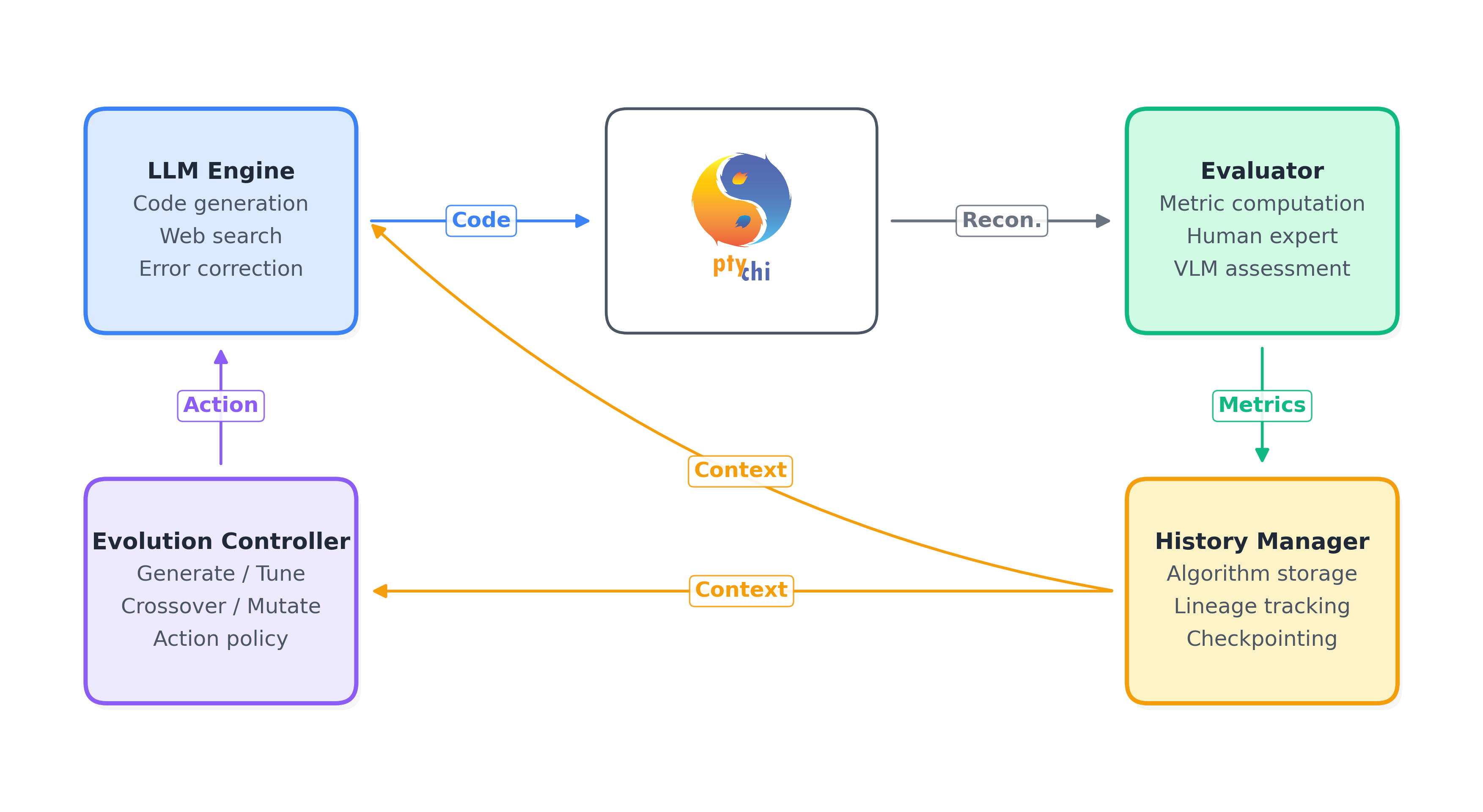}
\caption{System architecture of \ptychi{} showing the discovery loop. The LLM Engine generates regularizer code, which is executed by the Pty-Chi reconstruction library. The Evaluator assesses reconstruction quality and records metrics to the History Manager. Context from the history informs both the Evolution Controller (for selecting parent algorithms and actions) and the LLM Engine (for code generation). The Evolution Controller triggers actions (generate, tune, crossover, or mutate) that prompt the LLM to produce new regularizer variants.}
\label{fig:architecture}
\end{figure}

\ptychi{} is an autonomous framework for discovering regularization algorithms through LLM-guided evolutionary search. The system comprises four interconnected components: an LLM engine for algorithm generation and analysis, a multi-modal evaluation pipeline for assessing reconstruction quality, evolutionary refinement mechanisms for improving algorithms over generations, and a history management system for context management, interpretability and checkpointing. Figure~\ref{fig:architecture} provides an overview of the complete architecture.

\subsection{Problem Formulation}

Ptychographic reconstruction seeks to recover a complex-valued object function $\psi(\bfr)$ from diffraction intensity measurements. In the standard formulation, a coherent probe $P(\bfr)$ illuminates the object at a series of overlapping positions $\{\bfr_j\}$, producing exit waves that propagate to a detector where only intensity (not phase) is recorded. The forward model relates the object to measurements through:
\begin{equation}
I_j(\bfq) = \left| \mathcal{F}\left[ P(\bfr - \bfr_j) \cdot \psi(\bfr) \right] \right|^2 + \epsilon
\label{eq:forward}
\end{equation}
where $\mathcal{F}$ denotes the Fourier transform, $\bfq$ represents detector coordinates, and $\epsilon$ accounts for noise. For multislice ptychography, the object is represented as a stack of $S$ transmission functions $\{\psi_s\}_{s=1}^{S}$, with the exit wave computed through sequential propagation and multiplication operations.

Reconstruction algorithms iteratively minimize a data fidelity term while incorporating regularization to address ill-posedness:
\begin{equation}
\hat{\psi} = \arg\min_\psi \sum_j \mathcal{L}(I_j, \psi) + \lambda R(\psi)
\label{eq:reconstruction}
\end{equation}
Here, $\mathcal{L}$ measures agreement between predicted and observed intensities (typically amplitude-based or Gaussian likelihood), $R(\psi)$ encodes prior knowledge through regularization, and $\lambda$ controls the regularization strength. Our goal is to automatically discover effective forms of $R$ and associated optimization strategies.

\subsection{LLM-Driven Algorithm Generation}

The core of \ptychi{} is an LLM engine that generates regularization algorithms as executable Python functions. Each generated function is a method of Pty-Chi's \texttt{Object} class and follows a pre-defined standardized interface: \texttt{regularize\_llm(self)} operates on a complex-valued object tensor stored in \texttt{self.data} and writes the regularized result through \texttt{self.set\_data()}. This interface enables seamless integration with the Pty-Chi library's reconstruction pipeline, where the regularization function is called at each epoch.

The generation process begins with prompt construction. Each prompt includes several components that provide context for the LLM. The experimental context describes the imaging modality, sample characteristics, and expected challenges; for example, it may note that an integrated circuit sample exhibits grid artifacts from periodic structures, or that a multislice dataset requires handling of crosstalk between depth layers. The algorithm history summarizes previously attempted approaches with their performance metrics, enabling the LLM to learn from both successes and failures. When web search is enabled, the prompt incorporates relevant scientific literature retrieved through queries about regularization techniques, the specific imaging modality, and known challenges. Finally, technical requirements specify code constraints including the function signature, available imports (PyTorch, NumPy, SciPy), and interface conventions.

The LLM generates candidate algorithms in response to these prompts. Generated code undergoes multi-pass extraction to identify Python function definitions, handling variations in output formatting across different model responses. Extracted code is then subjected to security validation that detects potentially dangerous operations including system calls, file I/O, network access, and imports of restricted modules. In our implementation, validation combines an LLM-based security audit with a restricted execution environment that whitelists available built-ins and imports for the generated regularizer code.

When algorithms fail during evaluation due to syntax errors, runtime exceptions, or invalid outputs, \ptychi{} attempts automatic correction. The LLM receives the failed algorithm along with the error message and traceback, then generates a corrected version that addresses the specific failure. This correction mechanism enables recovery from common issues including tensor shape mismatches, numerical instabilities, and API misuse. Up to two correction attempts are permitted before marking an algorithm as failed, balancing recovery potential against computational cost.

\subsection{Multi-Modal Evaluation Pipeline}

Discovered algorithms must be evaluated to guide the evolutionary search process. Recognizing that different experimental contexts have different evaluation requirements, \ptychi{} supports three complementary evaluation modes.

Ground truth evaluation is applicable when reference images are available, either from simulation or from expensive high-quality reconstructions (with the goal to find a regularizer that can reduce the computation cost or to be more robust in terms of algorithmic parameters). The system computes standard image quality metrics including structural similarity index (SSIM), peak signal-to-noise ratio (PSNR), root mean squared error (RMSE), and mean absolute error (MAE). For multislice reconstructions, metrics are computed independently for each layer, then aggregated through configurable strategies including mean, median, minimum, or percentile-based combination. This per-layer analysis enables identification of regularizers that perform well across all depth slices versus those that sacrifice quality on certain layers.

Human expert evaluation enables deployment on novel samples where no ground truth exists. Domain experts view reconstructed images and provide quality assessments on a continuous zero-to-one scale, accompanied by qualitative feedback describing observed artifacts, resolution, and contrast. Experts may also suggest directions for subsequent algorithm development, which are incorporated into generation prompts. While more labor-intensive than automated metrics, human evaluation captures aspects of reconstruction quality that may not be reflected in simple numerical metrics.

Vision-language model (VLM) evaluation provides a middle ground that reduces expert burden while enabling assessment beyond ground truth metrics. In few-shot mode, the VLM learns from example images paired with quality scores, developing an implicit model of what constitutes good reconstruction for the specific imaging modality. In description mode, natural language criteria guide assessment, for example specifying that reconstructions should exhibit sharp edges, uniform background, and absence of ringing artifacts. Both modes produce structured evaluations compatible with the evolutionary search process. We refer our previous work \cite{umeike2025adapting} for readers interested in learning more about how we develop VLMs for ptychography tasks. 

Regardless of evaluation mode, each algorithm is executed by dynamically patching the \texttt{PlanarObject.regularize\_llm} method in the Pty-Chi library. Reconstruction proceeds with the discovered regularizer enabled, running for a fixed iteration count determined by the experimental configuration. The execution environment restricts available imports to a curated set (PyTorch, NumPy, SciPy, and their submodules) to ensure reproducibility and prevent unintended side effects.

\subsection{Evolutionary Refinement}

\ptychi{} take actions sequentially to navigate the algorithm design space, and there are three actions available: generation, tuning, and evolution. An LLM-driven and rules-regulated action policy determines which action to take based on the current state of the algorithm population.

Generation creates new algorithms from scratch, informed by the experimental context and history of prior attempts. The LLM synthesizes novel approaches by drawing on its knowledge of regularization techniques, adapting concepts from the literature to the specific imaging challenge at hand. Generation is the primary action during the warmup phase of discovery, establishing a diverse initial pool of regularizers before evolution or tuning begins.

Tuning refines high-performing algorithms by adjusting hyperparameters while preserving algorithmic structure. Given a successful algorithm, the LLM analyzes its components and proposes modified parameter values, such as regularization weights, iteration counts, or learning rates. Tuning is triggered when excellent algorithms exist but may benefit from fine-grained optimization, as indicated by performance that is good but not yet saturated.

Evolution represents the most distinctive mechanism in \ptychi{}, comprising two semantically-guided operations. Crossover takes two successful parent algorithms and generates an offspring that combines their complementary strengths. Unlike recombination in traditional genetic algorithms, the LLM understands what each parent algorithm does and intentionally merges their techniques. For instance, given one parent with an effective TV formulation and another with gradient correlation penalty for crosstalk mitigation, crossover might produce an offspring that incorporates both approaches within a unified optimization framework. Mutation takes a single parent and introduces variations, such as adding new components, removing redundant operations, or modifying mathematical formulations, while preserving core functionality that contributed to the parent's success.

The action policy balances exploration of new algorithmic ideas against exploitation of successful approaches. During the warmup phase (typically the first five to ten iterations), the policy focuses exclusively on generation to establish population diversity. Once sufficient successful algorithms exist, the policy begins selecting tuning and evolution actions based on configurable criteria. Tuning is triggered when excellent-tier algorithms exist and a specified number of iterations have passed since the last tuning action. Evolution is triggered when the population contains enough successful algorithms to enable meaningful crossover, again subject to iteration-based throttling. When the policy would select an action that cannot be executed (for instance, evolution when too few successful algorithms exist), it falls back to generation.

\subsection{History Management and Interpretability}

Evaluated algorithms are stored with comprehensive metadata enabling post-hoc analysis and interpretation. Each record includes the complete source code, generation number, action type (generated, tuned, or evolved), parent identities for evolved algorithms, evaluation metrics, and (when available) LLM-generated analysis. This analysis extracts the techniques employed by each algorithm (such as TV regularization, gradient correlation penalty, or adaptive weighting), identifies key parameters and their values, and suggests potential improvements based on observed performance.

Algorithms are classified into performance tiers based on configurable metric thresholds. For ground truth evaluation using SSIM, a typical configuration might classify algorithms achieving SSIM $\geq$ 0.90 as excellent, $\geq$ 0.80 as good, $\geq$ 0.60 as moderate, and below 0.60 as poor. These classifications inform action selection (tuning targets excellent algorithms, evolution requires successful algorithms) and history compression.

To bound memory usage while preserving information necessary for effective generation and evolution, the history undergoes periodic compression. The compression policy always retains the top-performing algorithms regardless of generation, ensuring that the best discoveries remain available. All excellent and good algorithms are preserved, as these represent the successful outcomes that justify the discovery effort. Among moderate and poor algorithms, the policy retains representative samples: the best within each tier and recently generated algorithms that provide context for subsequent generations. This selective retention maintains population diversity while preventing unbounded growth.

The discovery state is checkpointed periodically and upon interruption signals (SIGINT, SIGTERM), enabling resumption of long-running experiments. Checkpoints capture the current stored history, optimization statistics, and cached web search results needed to resume discovery. Upon restart, \ptychi{} loads the most recent checkpoint and continues discovery from where it left off.

The comprehensive history maintained by \ptychi{} enables interpretability analyses that would be impossible with black-box optimization. Algorithm lineage can be traced to understand which parent techniques contributed to successful offspring. Technique timelines reveal when specific innovations emerged and propagated through the population. Failure analysis identifies common pitfalls, including numerical instabilities, over-smoothing, and memory errors, that the discovery process learned to avoid. These analyses yield scientific insights about what makes regularization effective for specific imaging challenges.

\section{Experiments}

We evaluate \ptychi{} on three challenging ptychography datasets spanning different reconstruction challenges. These experiments demonstrate the framework's ability to discover effective regularizers across diverse conditions while providing insights into the discovered algorithms and evolution dynamics.

\subsection{Experimental Setup}

The evaluation encompasses three datasets chosen to represent distinct imaging modalities and challenges. The X-ray Integrated Circuit (IC) dataset consists of hard X-ray synchrotron measurements of a semiconductor integrated circuit sample. This dataset presents multiple challenges: the initial probe estimate is inaccurate, the scan step size is large relative to probe dimensions (causing aliasing), and the periodic metal line structures in the IC produce prominent grid artifacts in naive reconstructions. The combination of these factors makes regularization particularly important for achieving high-quality reconstructions.

The X-ray Multislice dataset is a simulated two-layer stack designed to evaluate handling of depth-dependent reconstruction. Layer 1 contains low-contrast, muted features while Layer 2 contains high-contrast, sharp edges. The primary challenge is cross-layer interference (crosstalk), where features from one depth erroneously appear in reconstructions of adjacent depths. This dataset tests the framework's ability to discover regularizers that enforce slice independence while preserving layer-specific characteristics.

The Electron Apoferritin dataset consists of low-dose electron ptychography simulations of apoferritin protein. The primary challenge is severe noise and reduced spatial resolution due to the dose limitations necessary to prevent radiation damage to biological specimens. This dataset evaluates the framework's ability to discover regularizers that suppress noise while preserving fine structural details.

For all datasets, we compare against reconstruction without regularization as implemented in the Pty-Chi library. This baseline represents standard ptychographic reconstruction using the least-square maximum likelihood algorithm (LSQML) \cite{Odstrcil2018-ns} without explicit priors beyond constraints inherent to the algorithm. We report improvements relative to this baseline to isolate the contribution of discovered regularizers.

The experiments uses OpenAI's O3 reasoning model for code generation, analysis and evolutionary operations. Web search is enabled for context retrieval, allowing the LLM to ground generation in scientific literature. During discovery, each reconstruction evaluation runs for a fixed iteration count determined by dataset characteristics: 1000 iterations for apoferritin, 1000 for multislice, and 3000 for the more challenging IC dataset. We report convergence analyses of the discovered algorithms in Figure~\ref{fig:convergence}.

\subsection{Main Results}

Table~\ref{tab:results} summarizes the discovery process and final performance across all three datasets. \ptychi{} successfully discovers regularizers that substantially outperform baselines on all datasets, with improvements ranging from +0.12 to +0.26 SSIM.

\begin{table}[htbp]
\caption{Discovery Statistics and Performance. Best reports the discovery-evaluation SSIM at the fixed discovery horizon (IC: 3000 iterations; apoferritin and multislice: 1000). $\Delta$ SSIM and $\Delta$ PSNR report improvement of the best discovered regularizer over the baseline at the representative comparison iteration in Figure~\ref{fig:convergence} (IC: 3000 iterations; apoferritin: 1000; multislice: 1000) using the same evaluation references as in Figure~\ref{fig:reconstruction}. For multislice, $\Delta$ values are computed on the mean across layers.}
\label{tab:results}
\begin{tabular}{lcccccc}
\toprule
Dataset & Algorithms & Success & Best & $\Delta$ SSIM & $\Delta$ PSNR & Duration \\
\midrule
X-ray IC & 100 & 83\% & 0.785 & +0.26 & +8.3 dB & 16.5 hrs \\
Apoferritin & 147 & 97\% & 0.836 & +0.12 & +3.2 dB & 29.5 hrs \\
Multislice & 97 & 94\% & 0.871 & +0.18 & +8.0 dB & 10.5 hrs \\
\bottomrule
\end{tabular}
\end{table}

The X-ray IC dataset exhibits the largest absolute improvement, with the best discovered regularizer achieving +0.26 SSIM and +8.3~dB PSNR over baseline at the discovery evaluation horizon (3000 iterations). This substantial gain reflects the severity of artifacts in naive reconstructions and the effectiveness of the discovered notch filtering approach for suppressing grid artifacts. The 83\% algorithm success rate, lower than other datasets, indicates that the challenging nature of this reconstruction problem leads to more failed attempts, though the framework successfully navigates these failures through its correction mechanism.

The apoferritin dataset shows more modest improvements (+0.12 SSIM, +3.2~dB PSNR) but achieves the highest success rate (97\%). The relatively smaller gains likely reflect fundamental limitations imposed by shot noise at low dose, as no regularizer can recover information that was never captured. Nevertheless, the discovered regularizers provide meaningful improvements in reconstruction quality that could impact downstream analysis.

The multislice dataset achieves the highest absolute reconstruction quality (0.871 SSIM), with mean improvement of +0.18 SSIM across layers (layer 1: +0.17; layer 2: +0.20). The 94\% success rate indicates that algorithms generated for this problem are more likely to produce valid reconstructions, potentially because the well-defined crosstalk challenge admits clearer algorithmic solutions. The discovery process found 4 excellent-tier and 30 good-tier algorithms, demonstrating that multiple effective approaches exist for this problem.

\begin{figure}[htbp]
\centering
\includegraphics[width=\columnwidth,height=0.72\textheight,keepaspectratio]{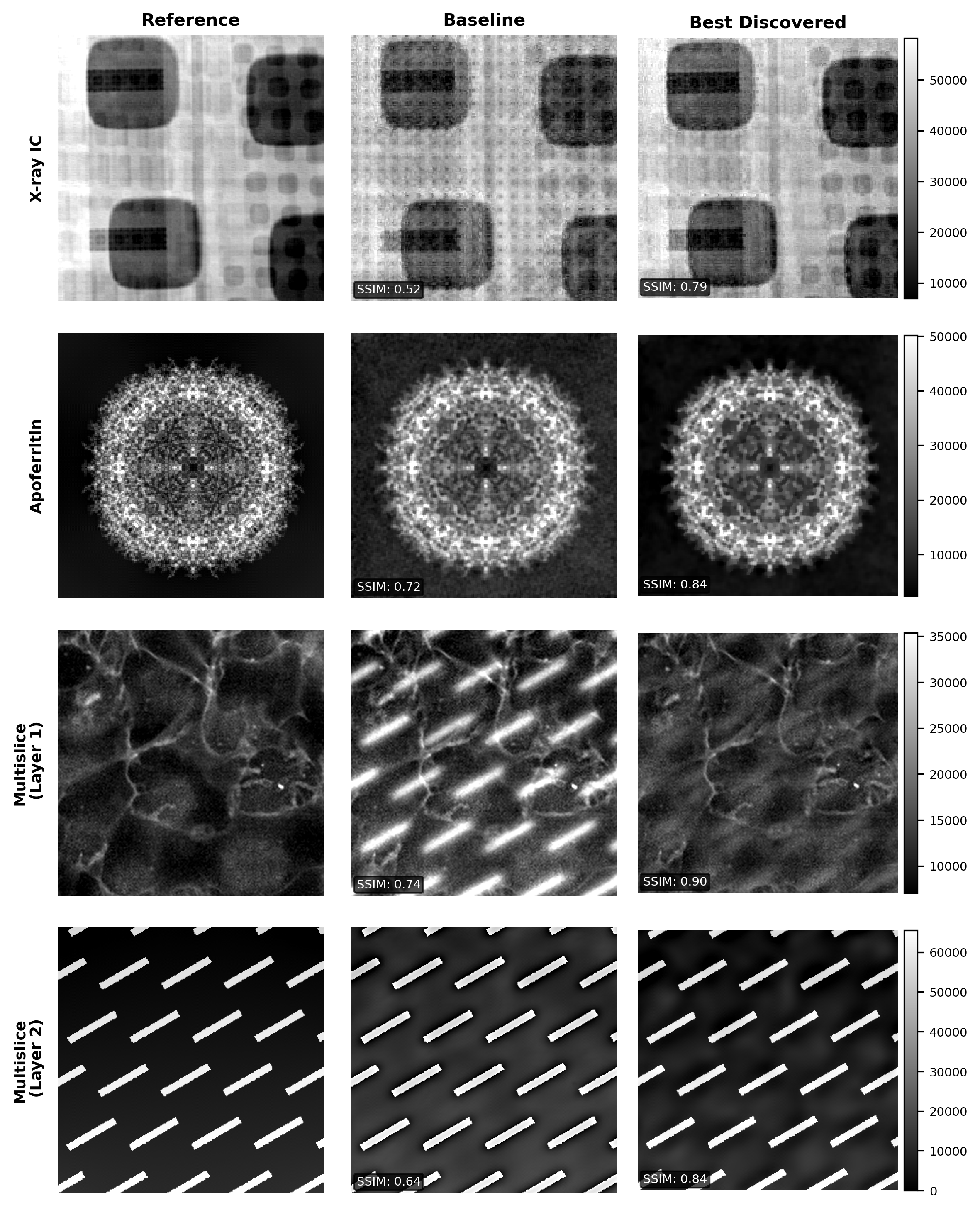}
\caption{Visual comparison of reconstructed phase images at representative iterations. Columns show a reference image, baseline reconstruction, and best discovered regularizer. For multislice and apoferritin, the reference is the simulated ground-truth phase; for X-ray IC, the reference is a long-iteration (i.e., 10,000) reconstruction used as an evaluation reference in the absence of ground truth. SSIM values overlaid on baseline/best are computed against the corresponding reference, and all images within each row share a unified color scale (shown by the colorbar on the right). X-ray IC is shown at 3000 iterations; apoferritin and multislice are shown at 1000 iterations.}
\label{fig:reconstruction}
\end{figure}

Figure~\ref{fig:reconstruction} provides visual comparison of reconstruction quality across datasets. The X-ray IC images reveal severe periodic artifacts in the baseline that are substantially reduced by the discovered notch filtering regularizer. For apoferritin, the improvement is more subtle but visible as reduced noise and sharper protein features. Both multislice layers show characteristic crosstalk artifacts (diagonal stripes from feature leakage between layers) in the baseline; layer 1 exhibits more complex structures while layer 2 shows cleaner diagonal features. The discovered gradient correlation penalty and Gram orthogonality techniques effectively suppress these artifacts in both layers.

\begin{figure}[htbp]
\centering
\begin{subfigure}[b]{0.48\columnwidth}
\includegraphics[width=\linewidth]{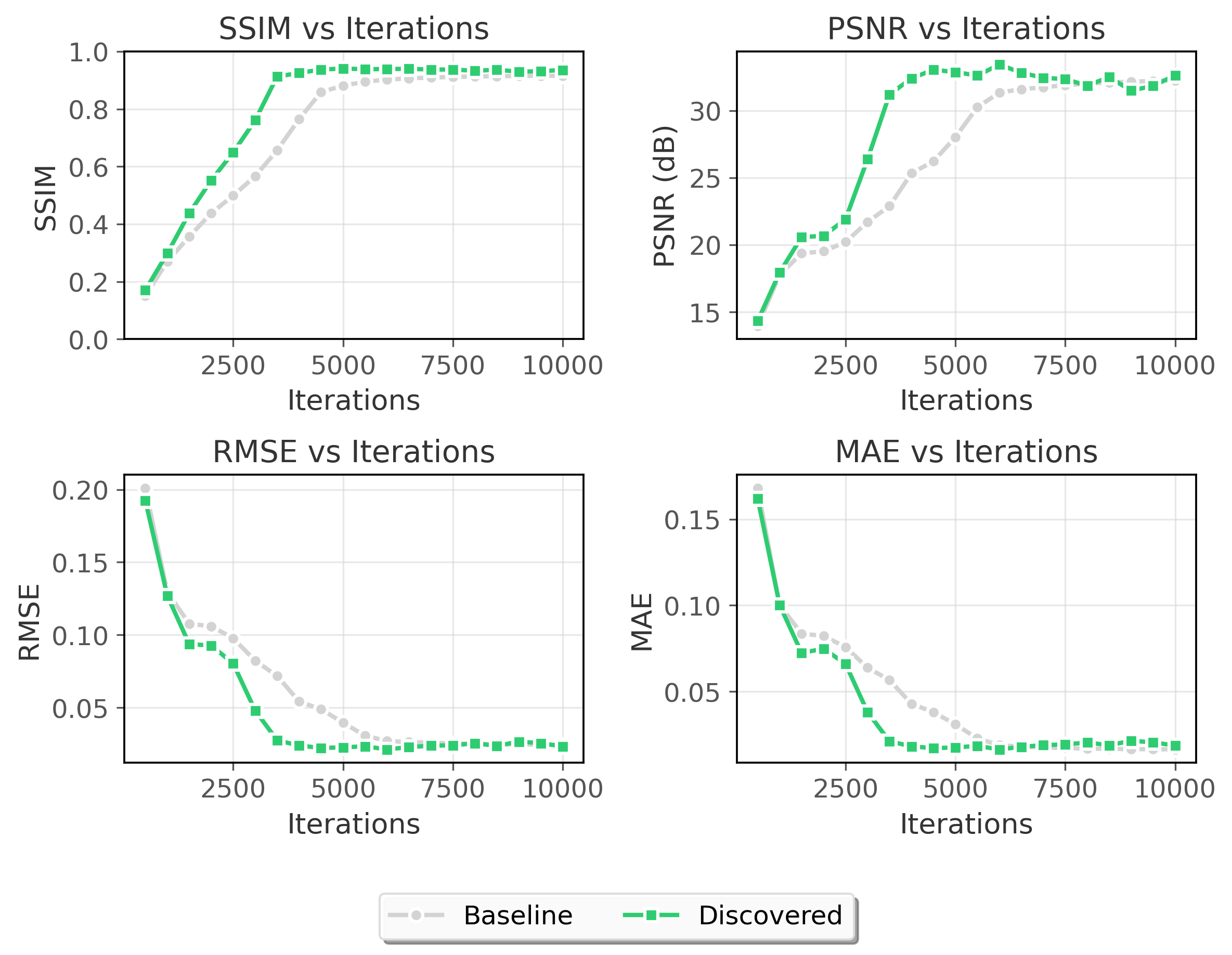}
\caption{X-ray IC}
\end{subfigure}
\hfill
\begin{subfigure}[b]{0.48\columnwidth}
\includegraphics[width=\linewidth]{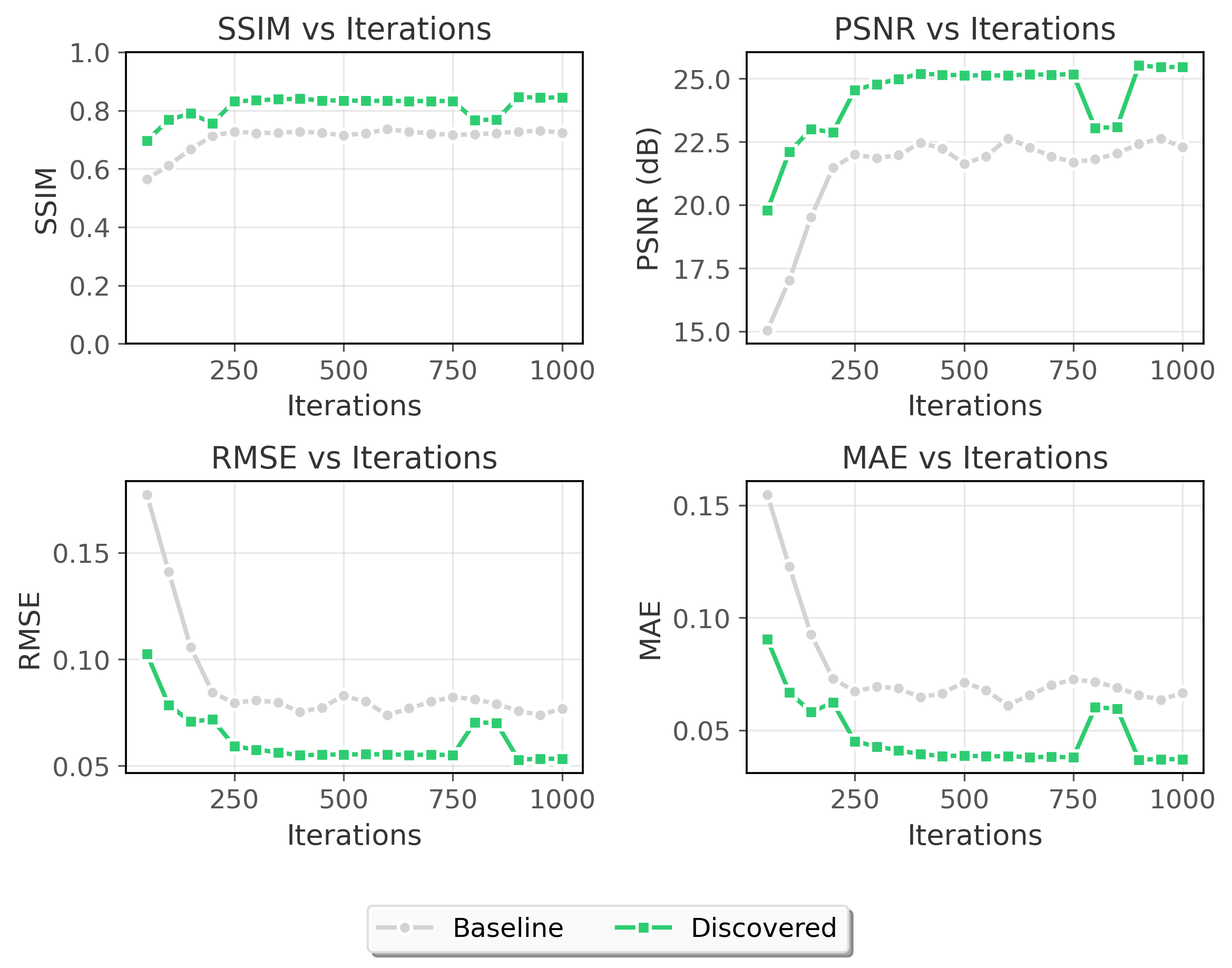}
\caption{Apoferritin}
\end{subfigure}

\begin{subfigure}[b]{\columnwidth}
\includegraphics[width=\linewidth]{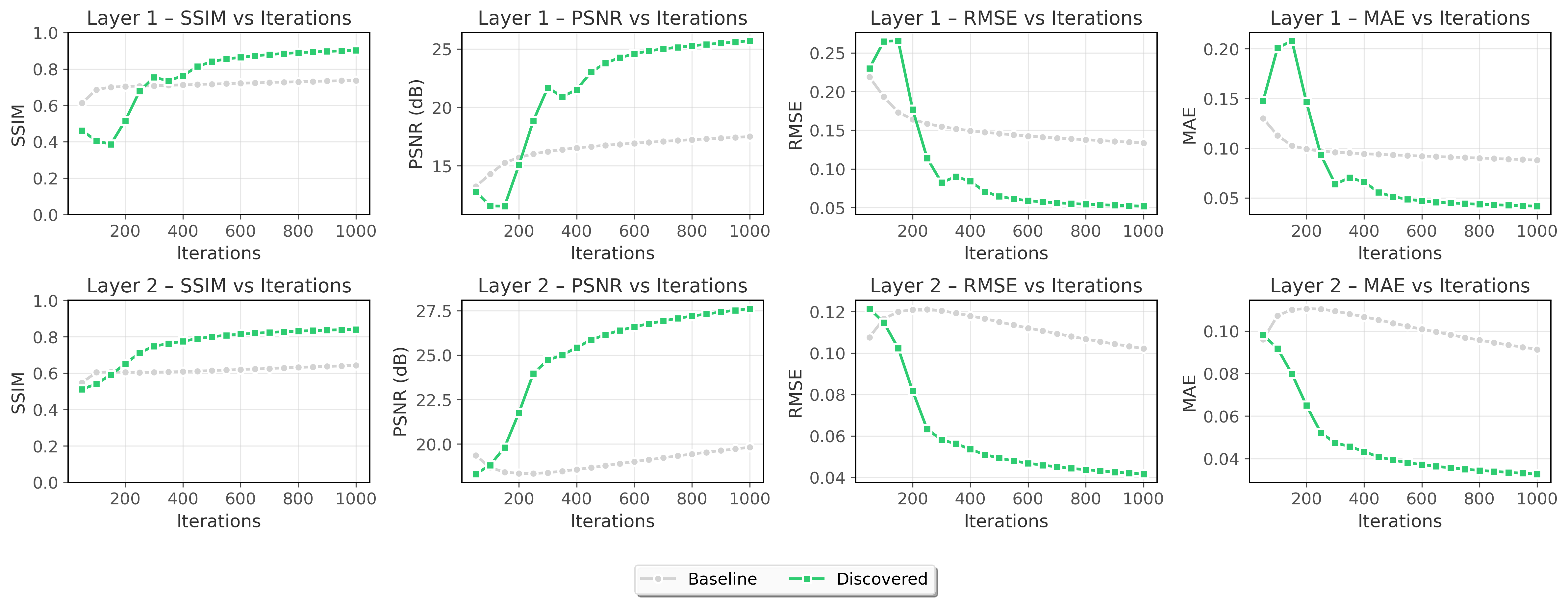}
\caption{Multislice}
\end{subfigure}
\caption{Convergence comparison showing SSIM, PSNR, RMSE, and MAE versus reconstruction iterations for baseline (grey) and best discovered regularizer (green), computed against the same evaluation references as in Figure~\ref{fig:reconstruction}. IC is swept up to 10,000 iterations; apoferritin and multislice are swept up to 1,000 iterations. Multislice results are shown per layer (top row: layer~1; bottom row: layer~2).}
\label{fig:convergence}
\end{figure}

Figure~\ref{fig:convergence} illustrates convergence dynamics across datasets. The X-ray IC results reveal substantial improvement at 3000 iterations, where baseline SSIM is 0.52 compared to 0.79 for the regularized reconstruction. At higher iteration counts, the baseline gradually catches up as increased computation partially compensates for the lack of explicit regularization, which in turn suggests that we can save time and computational resources with this discovered regularizer for similar IC datasets.

The multislice convergence curves reveal an interesting phenomenon: at early iterations (roughly the first 200-250), the regularized reconstruction can perform worse than baseline. This initial degradation occurs because the regularizer perturbs the reconstruction away from the data-consistent solution before its structure-enforcing properties become beneficial. By approximately 250-300 iterations, the regularized reconstruction overtakes baseline and maintains superior performance thereafter. This observation has practical implications: users should ensure sufficient iterations when deploying discovered regularizers.

\subsection{Analysis of Discovered Algorithms}

Examination of top-performing algorithms reveals dataset-specific innovations alongside common patterns. Table~\ref{tab:techniques} summarizes the key techniques discovered for each dataset.

\begin{table}[htbp]
\caption{Key Techniques Discovered by Dataset}
\label{tab:techniques}
\begin{tabular}{lll}
\toprule
Dataset & Technique & Description \\
\midrule
X-ray IC & Adaptive notch filtering & FFT-based grid artifact suppression \\
X-ray IC & Structure-tensor TV & Edge-aware anisotropic smoothing \\
X-ray IC & Barzilai-Borwein & Adaptive step size acceleration \\
Apoferritin & Variance-gated & Preserves features under noise \\
Apoferritin & Soft-Huber & Noise-adaptive shrinkage \\
Apoferritin & Perona-Malik & Anisotropic diffusion \\
Multislice & Gradient correlation penalty & Penalizes shared gradients \\
Multislice & Gram orthogonality & Slice independence via correlation \\
Multislice & Contrast-adaptive & Per-slice regularization strength \\
Multislice & Spectral masks & Butterworth filters per slice \\
\bottomrule
\end{tabular}
\end{table}

Several individual discoveries parallel established techniques in computational imaging, demonstrating that the LLM effectively retrieves and applies relevant domain knowledge. The notch filtering approach for grid artifact suppression has been demonstrated by Huang \emph{et al.},\cite{Huang2017-rz} Gram orthogonality for enforcing slice independence echoes multi-view geometry constraints used in stereo reconstruction, Perona-Malik diffusion is a classical edge-preserving filter,\cite{perona1990scale} and the Huber penalty provides a smooth approximation to TV that avoids non-differentiability at zero gradient.\cite{huber1964robust}

However, the LLM combines these known techniques in novel ways tailored to each challenge, producing algorithms whose overall design goes beyond what any single technique provides. Each best-performing algorithm integrates multiple complementary components with carefully orchestrated interaction between them, representing the kind of algorithmic synthesis that would be difficult to achieve through manual design or parameter tuning alone.

For example, the best regularizer discovered for the multislice case study integrates four gradient components within a single regularization step: (1)~isotropic 3-D Charbonnier-TV with reflective boundary conditions, where the TV penalty extends along the depth (inter-slice) direction in addition to the spatial dimensions; (2)~multi-slice gradient exclusion that penalizes locations where multiple slices simultaneously exhibit strong gradients, discouraging crosstalk; (3)~Gram orthogonality that minimizes off-diagonal entries of the slice-wise Gram matrix to enforce statistical independence between layers; and (4)~complementary Butterworth spectral masks, where low-contrast slices receive a high-pass filter and high-contrast slices receive a low-pass filter, with the filter strength exponentially annealed over epochs. A notable innovation is the contrast-adaptive TV weighting, which blends a reciprocal contrast map with a logistic contrast map (both percentile-clipped) to automatically assign stronger regularization to low-contrast slices and weaker regularization to high-contrast ones. The four gradient contributions are aggregated and passed through a two-stage robust clamp (scale-invariant median clamp followed by per-slice $\ell_2$ clamp) before being applied to the object. Rather than using a fixed step size, the regularizer applies this combined gradient using Adam update rules\cite{kingma2015adam} with persistent first- and second-moment estimates that accumulate across reconstruction epochs and a three-phase learning rate schedule (warm-up, exponential decay, and cosine annealing). After each update, per-slice $\ell_2$ renormalization prevents the regularization from altering the overall energy of any slice. The use of a stateful optimizer within a regularization step is unconventional: conventional regularizers typically apply a single closed-form operation per epoch (e.g., soft thresholding, one-step proximal TV). This design pattern was not prescribed by the framework and represents a novel strategy autonomously discovered by the LLM, enabling the regularizer to make larger, more effective updates in early epochs and converge smoothly as reconstruction progresses.

The top-performing X-ray IC algorithm takes the internal-optimization idea further. It decomposes the complex-valued object into amplitude and phase, then solves each component through a dedicated iterative sub-loop within every reconstruction epoch: 25 amplitude iterations, 15 phase iterations, and 2 joint complex-TV iterations. The amplitude path computes a structure tensor to estimate local edge orientation and applies anisotropic Huber-TV that penalizes gradients perpendicular to detected edges $4\times$ more strongly than gradients parallel to them, combined with a TGV-like second-order Laplacian term that reduces staircasing. The Huber threshold decays every four internal iterations (a form of continuation that progresses from smooth to sharp regularization). The phase path applies amplitude-weighted TV that smooths more aggressively in low-signal regions and wraps the result to $[-\pi, \pi)$. After the joint complex-TV step, FFT-based notch filtering suppresses periodic grid artifacts. Rather than requiring known artifact periodicity, the algorithm automatically detects the strongest peaks in the power spectrum (excluding the DC component), constructs symmetric Gaussian notch pairs at those locations, and suppresses them in Fourier space. Within the amplitude sub-loop, step sizes are adaptively selected using the Barzilai-Borwein (BB) method,\cite{barzilai1988two} which approximates the local curvature from successive gradient differences ($\tau_\mathrm{BB} = \mathbf{s} \cdot \mathbf{y} / \mathbf{y} \cdot \mathbf{y}$); a candidate step is accepted only when an Armijo-type energy decrease condition is satisfied, providing quasi-Newton acceleration with guaranteed descent. Positivity of the amplitude is enforced through a softplus transformation with a data-adaptive steepness parameter. The use of an iterative solver inside the regularization step effectively turns each epoch's regularization into a proximal operator that approximately solves the regularization sub-problem, a strategy that, like the Adam approach in multislice, was autonomously discovered rather than prescribed.

The best apoferritin algorithm employs a different strategy: rather than an optimization loop, it applies a sequential signal-processing pipeline that separately denoises the amplitude and corrects the phase. Amplitude processing begins with a noise-adaptive Soft-Huber shrinkage that pulls values toward unit amplitude, where the shrinkage threshold is gated by a sigmoid function of the global variance, yielding stronger denoising under noisier conditions and lighter denoising when the signal is already clean. This is followed by cascaded guided filters at two radii ($r{=}1$ and $r{=}2$), each blending the filtered output with the input in proportion to local variance, thereby preserving edges while suppressing noise in flat regions. An optional Gaussian fusion step activates when global noise exceeds a threshold. Phase processing consists of quality-guided Itoh unwrapping (row-then-column), followed by mean subtraction to remove any DC offset or residual ramp. The unwrapped phase is then smoothed by five steps of Perona-Malik diffusion\cite{perona1990scale} with a conduction coefficient $c = 1/(1 + |\nabla u|^2 / \lambda^2)$ that preserves edges while diffusing noise. Finally, the regularized object is formed by blending the denoised amplitude and corrected phase with the original estimate through a weak complex $L_2$ anchor ($\alpha = 0.1$), which prevents over-regularization and maintains fidelity to the data. This pipeline approach, while not involving an internal optimizer, demonstrates a different form of algorithmic sophistication: the careful ordering and interaction of multiple denoising and phase-correction stages, each conditioned on data statistics.

Looking across all three best algorithms, several cross-cutting patterns emerge. Certain techniques appear consistently regardless of dataset: edge-aware TV variants (Huber, Charbonnier, anisotropic) that preserve sharp features while smoothing noise; adaptive parameter scheduling that reduces regularization strength over iterations (spectral annealing in multislice, Huber continuation in IC, variance-gated thresholds in apoferritin); and robust gradient handling including clipping, normalization, and adaptive step sizes. These patterns suggest general principles of effective regularization that transfer across imaging modalities. Two structural patterns are also noteworthy. First, a polar decomposition strategy is independently adopted by both the IC and apoferritin algorithms: each decomposes the complex-valued object into amplitude and phase, applies tailored regularization to each component, and recombines. This separation allows the regularizer to exploit domain-specific priors (e.g., amplitude positivity, phase continuity) that would be difficult to enforce on the complex-valued representation directly. Second, two distinct regularization architectures coexist: the IC and multislice algorithms embed internal optimization loops within the regularization step (iterative sub-loops or stateful Adam updates), while the apoferritin algorithm uses a sequential signal-processing pipeline. Both approaches are effective, suggesting that the optimal regularization architecture depends on the problem structure.

Dataset-specific innovations further illustrate the LLM's ability to reason about problem structure. The automatic notch filtering for IC grid artifacts detects artifact frequencies directly from the power spectrum rather than requiring user specification, demonstrating the framework's ability to discover self-calibrating solutions. The gradient correlation penalty and Gram orthogonality for multislice address slice ambiguity from complementary perspectives: gradient exclusion acts locally (penalizing shared edges), while Gram orthogonality acts globally (enforcing statistical independence), and their combination via crossover proved more effective than either alone. The noise-adaptive shrinkage and variance-gated filtering for apoferritin specifically target the signal-dependent noise characteristics of low-dose electron imaging.

\subsection{Evolution Dynamics}

The discovery process exhibits distinct dynamics across datasets, reflecting differences in problem structure and the effectiveness of evolutionary operations. 
For the multislice and apoferritin datasets, crossover operations produced the majority of top-performing algorithms. This pattern suggests that these problems admit compositional solutions where distinct algorithmic components address orthogonal aspects of the challenge. For example, the best multislice regularizer was produced through crossover of two successful parents, combining one parent's TV formulation with another's slice independence constraints. The LLM's ability to understand what each parent does, rather than blindly recombining code fragments, enables meaningful synthesis.

The X-ray IC dataset presents a contrasting pattern: all top-five algorithms were freshly generated rather than evolved through crossover or mutation. We hypothesize that the unique challenge of grid artifacts requires novel solutions (specifically, notch filtering in the frequency domain) that were not present in the initial population and could not be obtained by recombining existing approaches. Fresh generation, informed by the experimental context describing periodic structures, proved more effective than evolutionary refinement for this problem.

\begin{figure}[htbp]
\centering
\begin{subfigure}[b]{0.48\columnwidth}
\includegraphics[width=\linewidth]{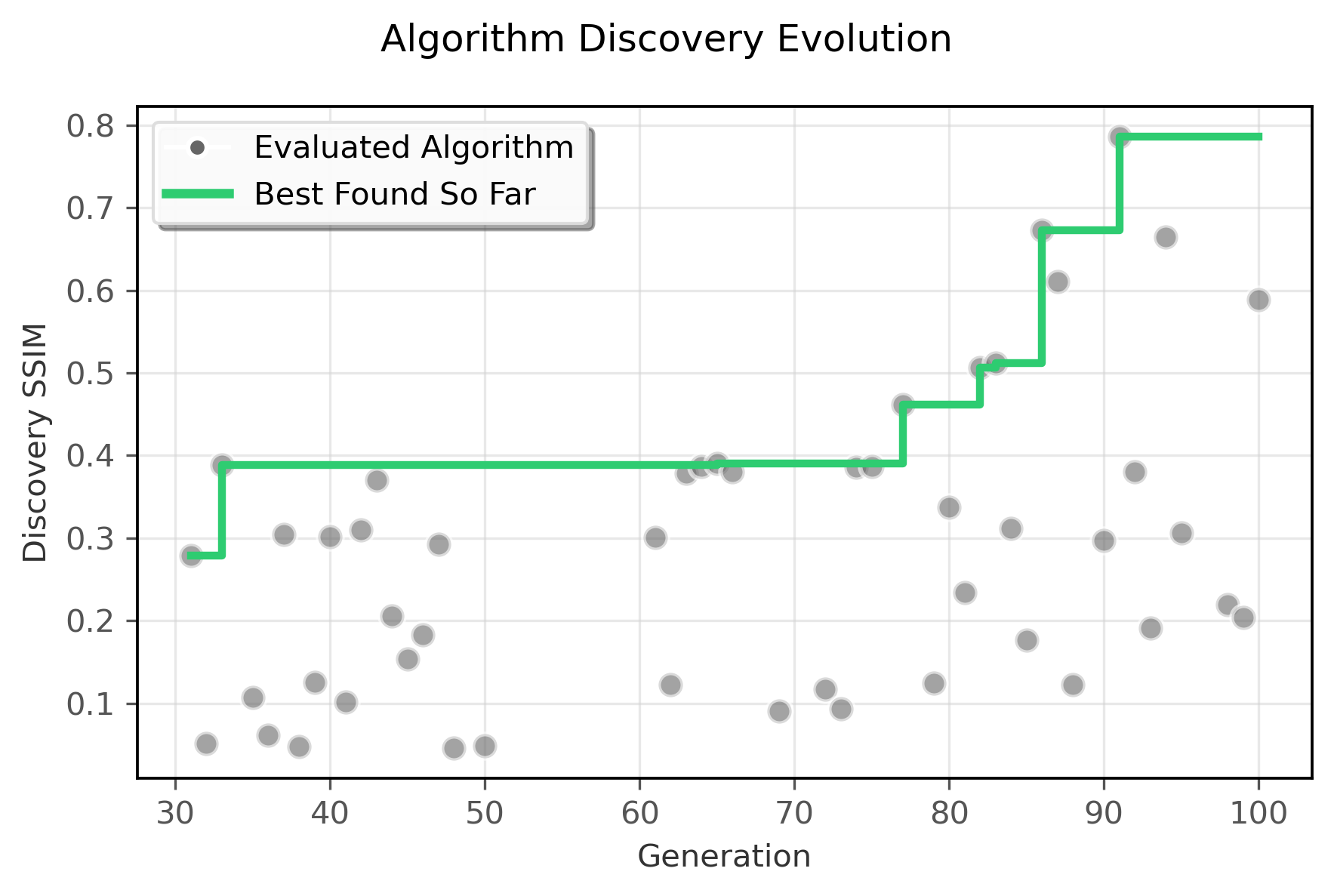}
\caption{X-ray IC}
\end{subfigure}
\hfill
\begin{subfigure}[b]{0.48\columnwidth}
\includegraphics[width=\linewidth]{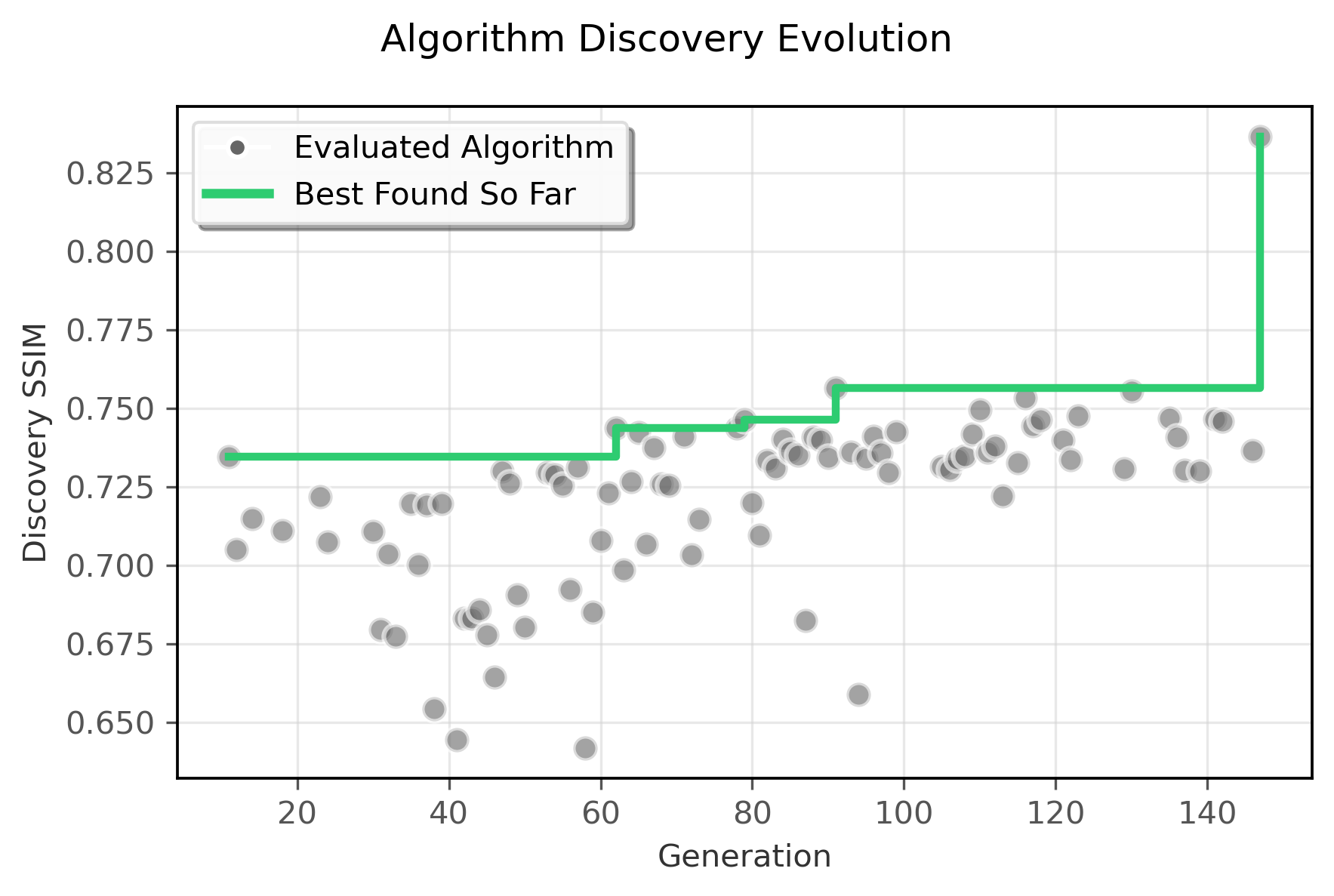}
\caption{Apoferritin}
\end{subfigure}

\begin{subfigure}[b]{0.48\columnwidth}
\includegraphics[width=\linewidth]{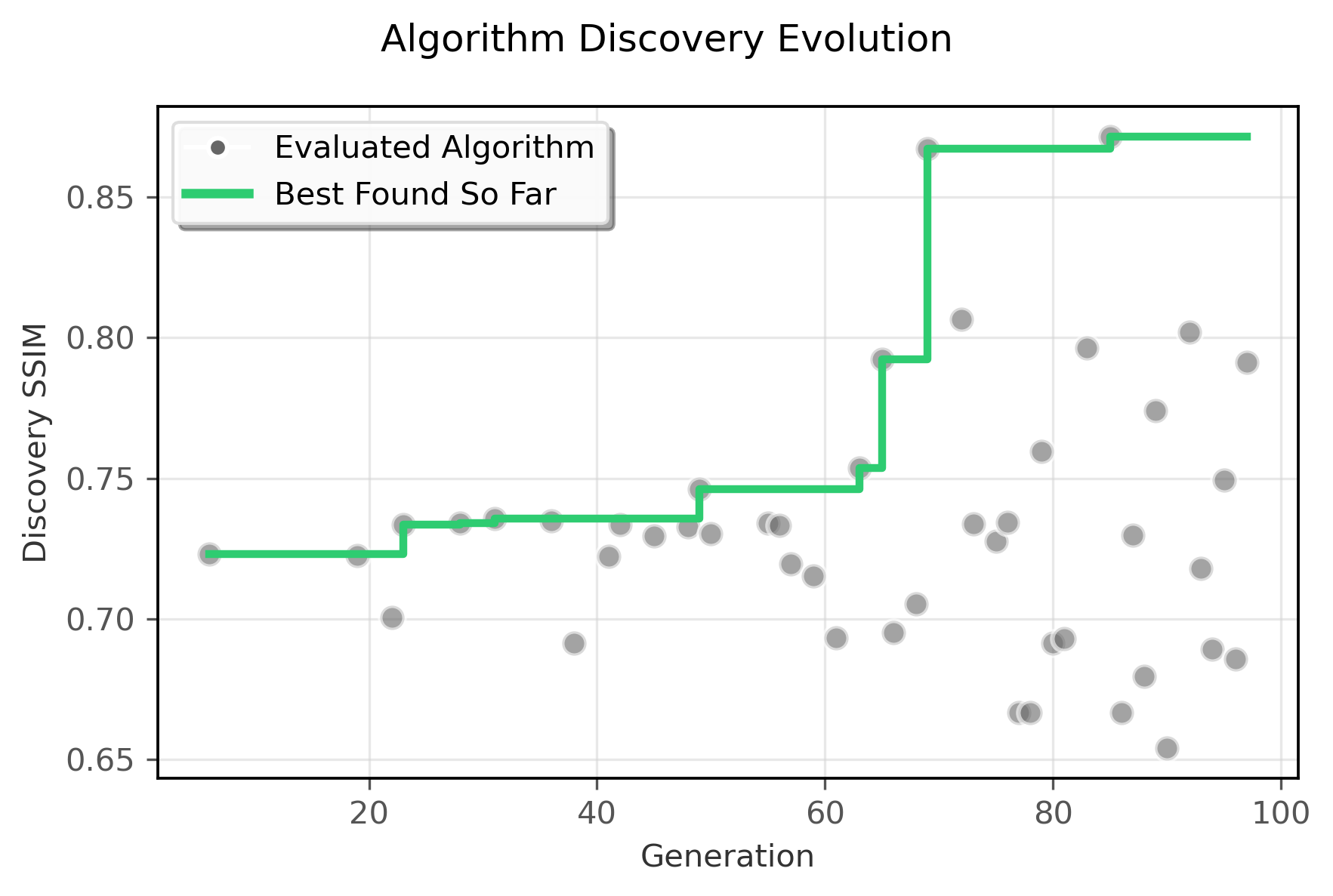}
\caption{Multislice}
\end{subfigure}
\caption{Discovery trajectory over generation number from the merged history. Grey points show successfully evaluated algorithms; the green curve shows the best discovery-evaluation SSIM found so far (computed at the fixed discovery horizon: IC 3000 iterations; apoferritin and multislice 1000).}
\label{fig:trajectory}
\end{figure}

Figure~\ref{fig:trajectory} visualizes the discovery trajectories. Performance generally improves over generations as the history accumulates successful algorithms that inform subsequent generation and evolution. For multislice, a clear inflection point occurs around generation 69 when crossover first combines gradient correlation penalty with Gram orthogonality, producing a large jump in SSIM that then stabilizes through subsequent generations. For X-ray IC, improvement is more gradual as the framework explores diverse approaches before converging on notch filtering strategies.

\begin{figure}[htbp]
\centering
\begin{subfigure}[b]{0.56\columnwidth}
\includegraphics[width=\linewidth]{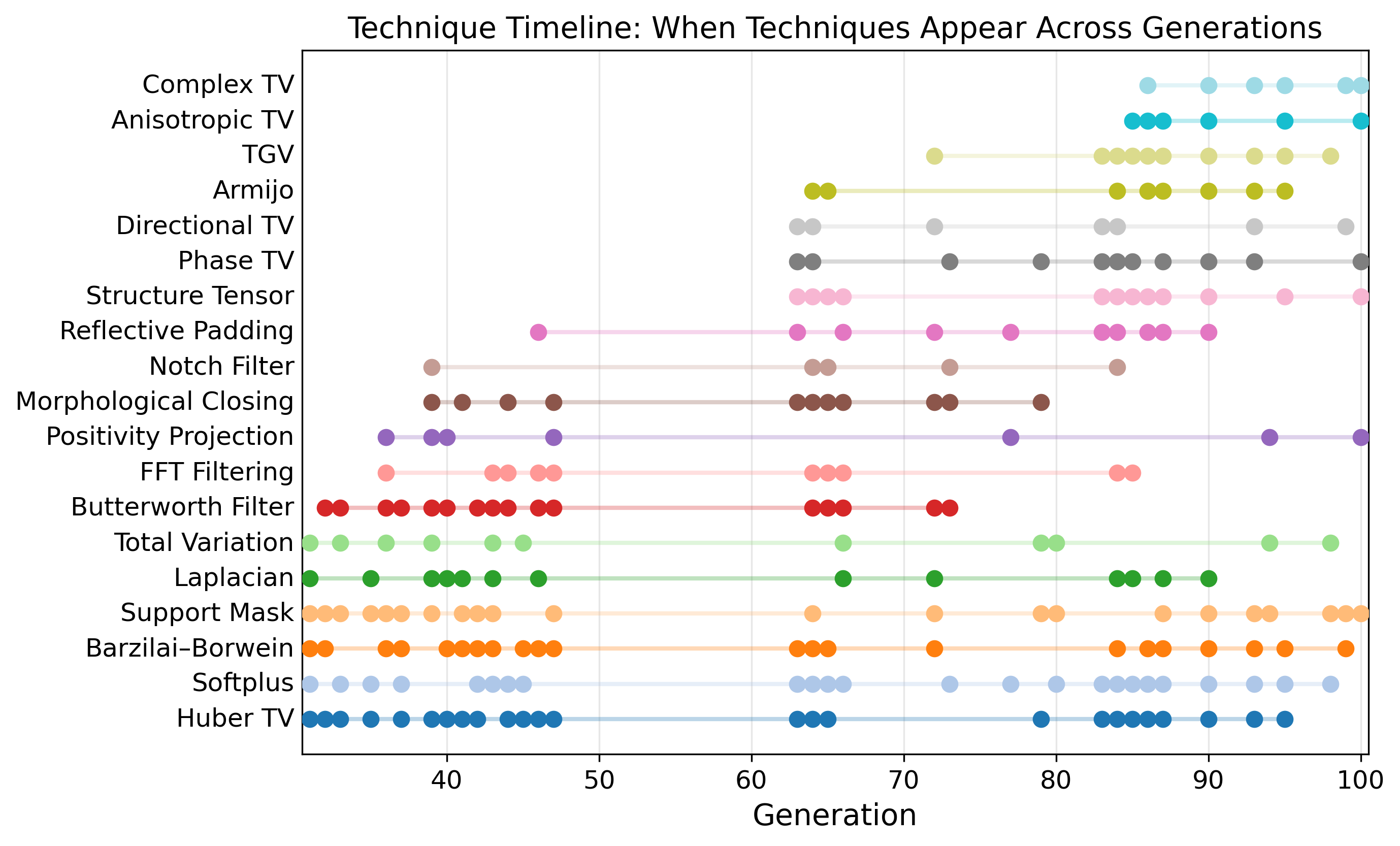}
\caption{X-ray IC}
\end{subfigure}
\hfill
\begin{subfigure}[b]{0.56\columnwidth}
\includegraphics[width=\linewidth]{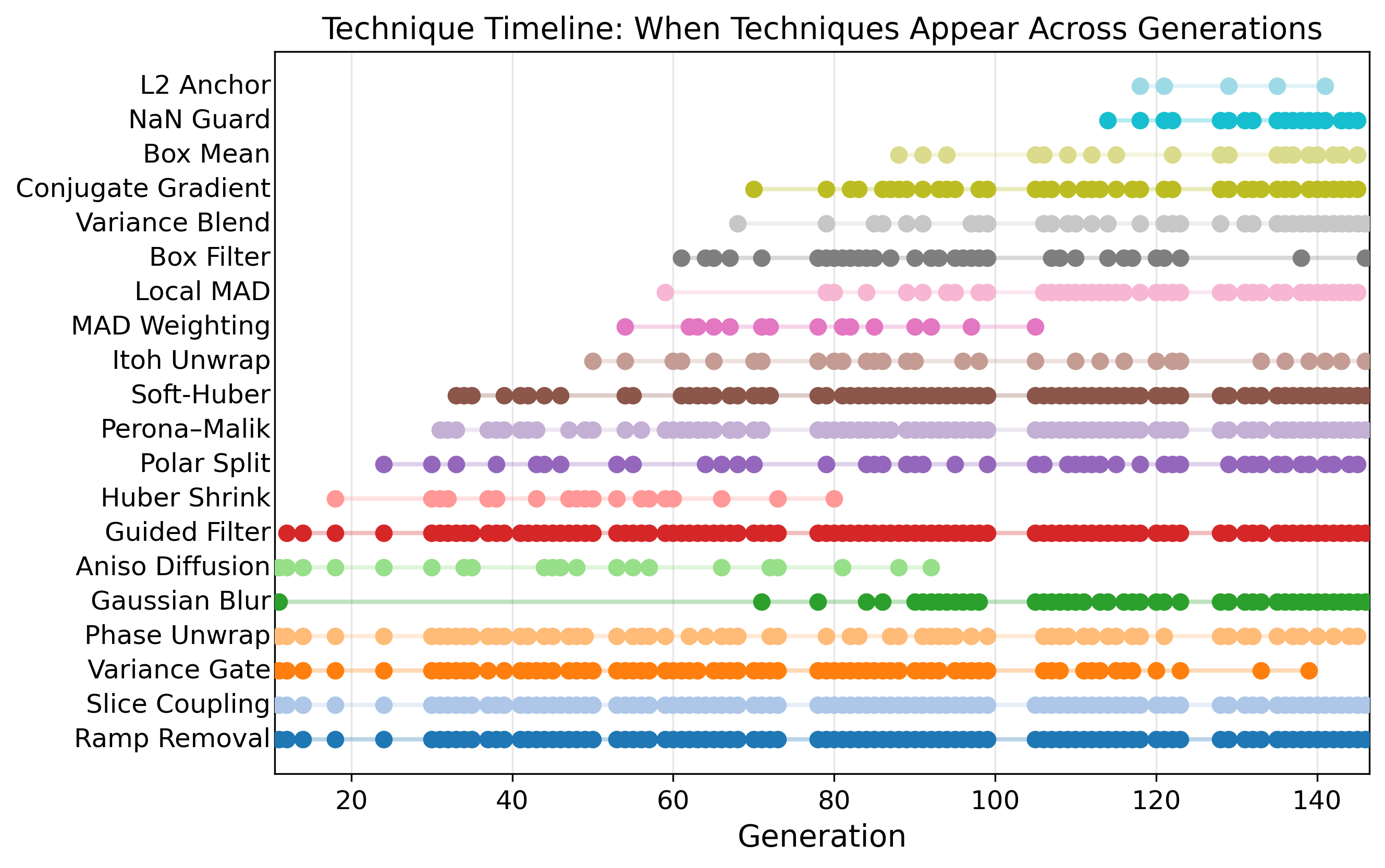}
\caption{Apoferritin}
\end{subfigure}

\begin{subfigure}[b]{0.56\columnwidth}
\includegraphics[width=\linewidth]{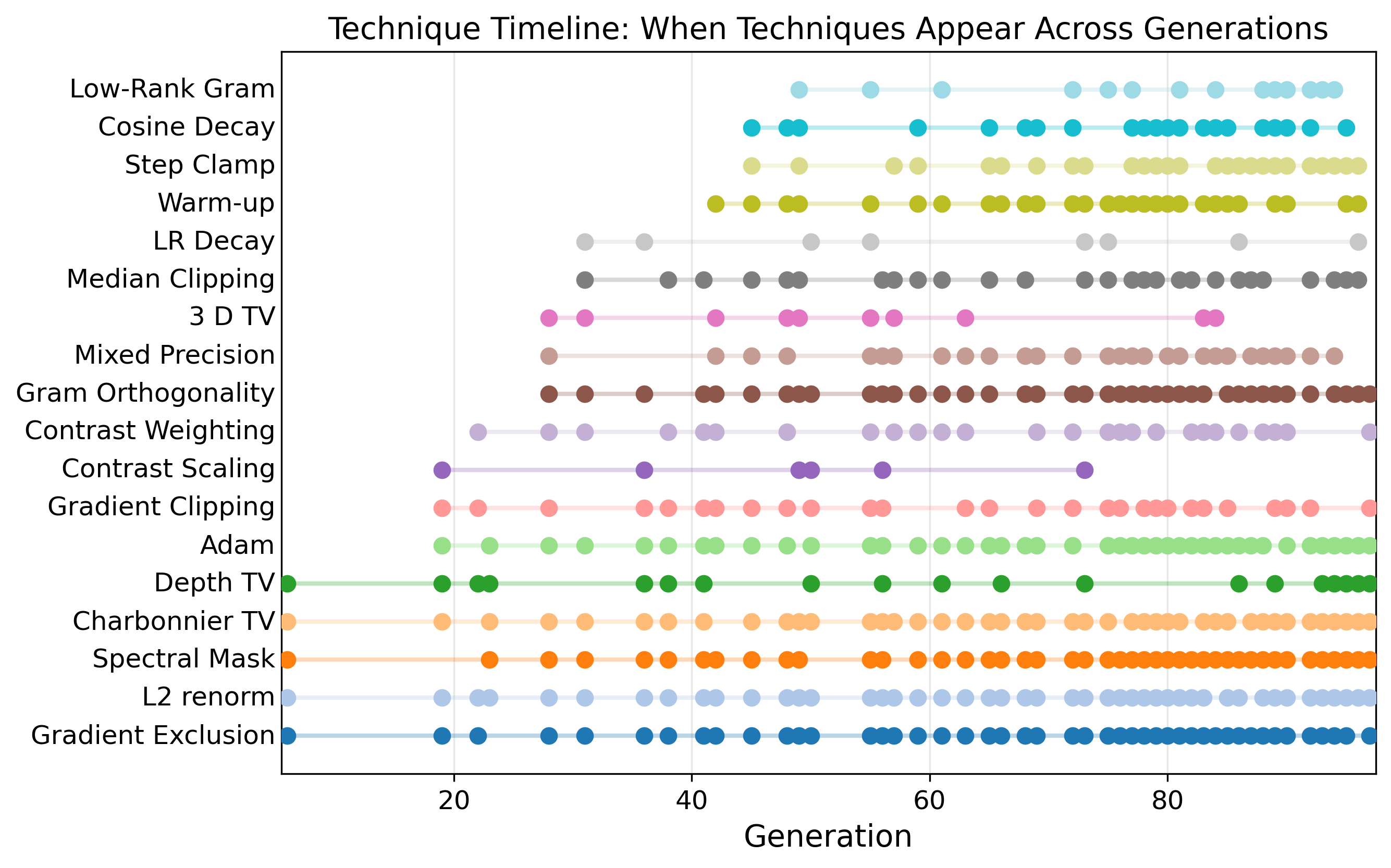}
\caption{Multislice}
\end{subfigure}
\caption{Technique adoption timeline from the merged history, showing the generations in which successfully evaluated algorithms include each technique. Lines span first-to-last appearance and markers indicate that a technique appears in at least one algorithm in that generation (techniques shown have $\geq$5 occurrences; technique names are normalized to merge spelling variants).}
\label{fig:timeline}
\end{figure}

The technique timeline analysis (Figure~\ref{fig:timeline}) reveals the emergence and propagation of algorithmic innovations. For the multislice dataset, gradient correlation penalty appears early (generation 6), Gram orthogonality appears later (generation 28), and their combination via crossover yields the best algorithms by generation 85. This pattern demonstrates that the evolutionary process can discover complementary techniques and effectively combine them.

\subsection{Limitations and Future Directions}

Several limitations of the current work suggest directions for future research. The computational cost of discovery remains substantial, requiring 10-30 hours per dataset due to the time required for reconstruction evaluation (approximately 10 minutes per algorithm) and API latency for LLM operations. Parallelizing evaluations and exploring more efficient reconstruction methods (e.g., early stopping) could substantially accelerate discovery.

The evaluation for the three case studies relies primarily on ground truth metrics computed against reference images. While we support human and VLM evaluation modes, our experiments use simulated or well-characterized samples where ground truth is available for objective evaluation and comparison. We intend to demonstrate deploying \ptychi{} on novel samples without references with human assessment for the next step.

The discovered algorithms are optimized for specific datasets and imaging conditions, which suggestes limited direct transferability: a regularizer discovered for one sample may not be optimal for a different sample with different feature scales or artifact characteristics. Re-running discovery for each new experimental condition is currently necessary, though this cost may be acceptable for repeated imaging of similar samples (e.g., IC) or high-value measurements where reconstruction quality is critical.

The search space is constrained by the requirement to interface with Pty-Chi's API, which prescribes the function signature and available state and importable packages. More general frameworks that allow modification of the iteration structure itself, not just the regularization function, could enable discovery of fundamentally new reconstruction algorithms beyond regularization.

Future directions include multi-objective discovery that optimizes for both reconstruction quality and computational efficiency, enabling deployment in time-sensitive applications. Transfer learning approaches could bootstrap discovery on new datasets using algorithms discovered for related problems. Integration with learned priors, such as deep image prior\cite{ulyanov2018dip} or implicit neural representations, which share a similar iterative workflow with conventional algorithms but represent the object with a neural network, could combine the interpretability of discovered regularization strategies with the representational power of deep learning.

\section{Conclusion}

We have presented \ptychi{}, an autonomous framework for discovering regularization algorithms for ptychographic reconstruction through LLM-guided evolutionary search. By combining large language model capabilities in code generation, scientific reasoning, and iterative refinement with systematic evaluation and evolutionary mechanisms, \ptychi{} can discover algorithms that outperform conventional approaches.

This work demonstrates that LLMs can effectively navigate the space of regularization algorithms, generating valid executable code that implements novel strategies informed by scientific domain knowledge. Evolutionary operations enable compositional discoveries that combine complementary techniques, with the LLM's semantic understanding enabling meaningful synthesis rather than random recombination. The discovered algorithms are interpretable, yielding scientific insights about effective regularization strategies for different imaging challenges. Analysis of evolution histories reveals patterns in technique emergence and propagation that characterize the discovery process.

This work represents a novel direction for AI-driven algorithm discovery in scientific computing. The approach can be generalized beyond ptychography to other inverse problems in imaging, such as magnetic resonance imaging, computed tomography, and seismic imaging, wherever handcrafted regularization represents a bottleneck to automation and optimization. More broadly, the combination of LLM-guided search with systematic evaluation represents a powerful paradigm for exploring algorithm design spaces that are too large and complex for manual exploration.

Finally, \ptychi{} advances the vision of autonomous scientific discovery by showing that large language models can not only assist human researchers but independently discover algorithmic innovations that enhance our ability to image the nanoscale world. As LLM capabilities continue to advance and computational resources become more accessible, we anticipate that autonomous algorithm discovery will become an increasingly important tool in the scientific toolkit.

\begin{acknowledgement}
This material is based upon work supported by Laboratory Directed Research and Development (LDRD) funding from Argonne National Laboratory (Award 2023-0049), provided by the Director, Office of Science, of the U.S. Department of Energy under Contract No. DE-AC02-06CH11357. This research used resources of the Advanced Photon Source, a U.S. Department of Energy (DOE) Office of Science User Facility operated for the DOE Office of Science by Argonne National Laboratory under Contract No. DE-AC02-06CH11357.
\end{acknowledgement}

\bibliography{references}

@article{lu2024aiscientist,
  title={The ai scientist: Towards fully automated open-ended scientific discovery},
  author={Lu, Chris and Lu, Cong and Lange, Robert Tjarko and Foerster, Jakob and Clune, Jeff and Ha, David},
  journal={arXiv preprint arXiv:2408.06292},
  year={2024}
}

@article{romeraparedes2024funsearch,
  title={Mathematical discoveries from program search with large language models},
  author={Romera-Paredes, Bernardino and Barekatain, Mohammadamin and Novikov, Alexander and Balog, Matej and Kumar, M Pawan and Dupont, Emilien and Ruiz, Francisco JR and Ellenberg, Jordan S and Wang, Pengming and Fawzi, Omar and others},
  journal={Nature},
  volume={625},
  number={7995},
  pages={468--475},
  year={2024},
  publisher={Nature Publishing Group UK London}
}

@article{novikov2025alphaevolve,
  title={AlphaEvolve: A coding agent for scientific and algorithmic discovery},
  author={Novikov, Alexander and V{\~u}, Ng{\^a}n and Eisenberger, Marvin and Dupont, Emilien and Huang, Po-Sen and Wagner, Adam Zsolt and Shirobokov, Sergey and Kozlovskii, Borislav and Ruiz, Francisco JR and Mehrabian, Abbas and others},
  journal={arXiv preprint arXiv:2506.13131},
  year={2025}
}

@article{thibault2008highres,
  title={High-resolution scanning x-ray diffraction microscopy},
  author={Thibault, Pierre and Dierolf, Martin and Menzel, Andreas and Bunk, Oliver and David, Christian and Pfeiffer, Franz},
  journal={Science},
  volume={321},
  number={5887},
  pages={379--382},
  year={2008},
  publisher={American Association for the Advancement of Science}
}

@article{pfeiffer2018xray,
  title={X-ray ptychography},
  author={Pfeiffer, Franz},
  journal={Nature Photonics},
  volume={12},
  number={1},
  pages={9--17},
  year={2018},
  publisher={Nature Publishing Group UK London}
}

@article{guizarsicairos2021ptychography,
  title={Ptychography: A solution to the phase problem},
  author={Guizar-Sicairos, Manuel and Thibault, Pierre},
  journal={Physics Today},
  volume={74},
  number={9},
  pages={42--48},
  year={2021},
  publisher={AIP Publishing}
}

@article{chen2021electron,
  title={Electron ptychography achieves atomic-resolution limits set by lattice vibrations},
  author={Chen, Zhen and Jiang, Yi and Shao, Yu-Tsun and Holtz, Megan E and Odstr{\v{c}}il, Michal and Guizar-Sicairos, Manuel and Hanke, Isabelle and Ganschow, Steffen and Schlom, Darrell G and Muller, David A},
  journal={Science},
  volume={372},
  number={6544},
  pages={826--831},
  year={2021},
  publisher={American Association for the Advancement of Science}
}

@article{marchesini2007unified,
  title={Invited article: A unified evaluation of iterative projection algorithms for phase retrieval},
  author={Marchesini, Stefano},
  journal={Review of scientific instruments},
  volume={78},
  number={1},
  year={2007},
  publisher={AIP Publishing}
}

@article{rudin1992tv,
  title={Nonlinear total variation based noise removal algorithms},
  author={Rudin, Leonid I and Osher, Stanley and Fatemi, Emad},
  journal={Physica D: nonlinear phenomena},
  volume={60},
  number={1-4},
  pages={259--268},
  year={1992},
  publisher={Elsevier}
}

@article{bredies2010tgv,
  title={Total generalized variation},
  author={Bredies, Kristian and Kunisch, Karl and Pock, Thomas},
  journal={SIAM Journal on Imaging Sciences},
  volume={3},
  number={3},
  pages={492--526},
  year={2010},
  publisher={SIAM}
}

@book{weickert1998anisotropic,
  title={Anisotropic diffusion in image processing},
  author={Weickert, Joachim and others},
  volume={1},
  year={1998},
  publisher={Teubner Stuttgart}
}

@inproceedings{ulyanov2018dip,
  title={Deep image prior},
  author={Ulyanov, Dmitry and Vedaldi, Andrea and Lempitsky, Victor},
  booktitle={Proceedings of the IEEE conference on computer vision and pattern recognition},
  pages={9446--9454},
  year={2018}
}

@inproceedings{venkatakrishnan2013pnp,
  title={Plug-and-play priors for model based reconstruction},
  author={Venkatakrishnan, Singanallur V and Bouman, Charles A and Wohlberg, Brendt},
  booktitle={2013 IEEE global conference on signal and information processing},
  pages={945--948},
  year={2013},
  organization={IEEE}
}

@article{zhang2021pnp,
  title={Plug-and-play image restoration with deep denoiser prior},
  author={Zhang, Kai and Li, Yawei and Zuo, Wangmeng and Zhang, Lei and Van Gool, Luc and Timofte, Radu},
  journal={IEEE Transactions on Pattern Analysis and Machine Intelligence},
  volume={44},
  number={10},
  pages={6360--6376},
  year={2021},
  publisher={IEEE}
}

@inproceedings{cam2025diffusion,
  title={Ptychographic image reconstruction from limited data via score-based diffusion models with physics-guidance},
  author={Cam, Refik M and Deng, Junjing and Kettimuthu, Rajkumar and Cherukara, Mathew J and Bicer, Tekin},
  booktitle={2025 IEEE 35th International Workshop on Machine Learning for Signal Processing (MLSP)},
  pages={1--6},
  year={2025},
  organization={IEEE}
}

@article{du2021ddip,
  title={Using a modified double deep image prior for crosstalk mitigation in multislice ptychography},
  author={Du, Ming and Huang, Xiaojing and Jacobsen, Chris},
  journal={Journal of Synchrotron Radiation},
  volume={28},
  number={4},
  pages={1137--1145},
  year={2021},
  publisher={International Union of Crystallography}
}

@article{yin2024pearfull,
  title={PEAR: a robust and flexible automation framework for ptychography enabled by multiple large language model agents},
  author={Yin, Xiangyu and Shi, Chuqiao and Han, Yimo and Jiang, Yi},
  journal={arXiv preprint arXiv:2410.09034},
  year={2024}
}

@article{liu2024eoh,
  title={Evolution of heuristics: Towards efficient automatic algorithm design using large language model},
  author={Liu, Fei and Tong, Xialiang and Yuan, Mingxuan and Lin, Xi and Luo, Fu and Wang, Zhenkun and Lu, Zhichao and Zhang, Qingfu},
  journal={arXiv preprint arXiv:2401.02051},
  year={2024}
}

@article{vanstein2024llamea,
  title={Llamea: A large language model evolutionary algorithm for automatically generating metaheuristics},
  author={van Stein, Niki and B{\"a}ck, Thomas},
  journal={IEEE Transactions on Evolutionary Computation},
  year={2024},
  publisher={IEEE}
}

@article{zimmermann2025llm,
  title={34 Examples of LLM Applications in Materials Science and Chemistry: Towards Automation, Assistants, Agents, and Accelerated Scientific Discovery},
  author={Zimmermann, Yoel and Bazgir, Adib and Al-Feghali, Alexander and Ansari, Mehrad and Bocarsly, Joshua and Brinson, L Catherine and Chiang, Yuan and Circi, Defne and Chiu, Min-Hsueh and Daelman, Nathan and others},
  journal={arXiv preprint arXiv:2505.03049},
  year={2025}
}

@article{prince2024opportunities,
  title={Opportunities for retrieval and tool augmented large language models in scientific facilities},
  author={Prince, Michael H and Chan, Henry and Vriza, Aikaterini and Zhou, Tao and Sastry, Varuni K and Luo, Yanqi and Dearing, Matthew T and Harder, Ross J and Vasudevan, Rama K and Cherukara, Mathew J},
  journal={npj Computational Materials},
  volume={10},
  number={1},
  pages={251},
  year={2024},
  publisher={Nature Publishing Group UK London}
}

@article{kamilov2023pnp,
  title={Plug-and-play methods for integrating physical and learned models in computational imaging: Theory, algorithms, and applications},
  author={Kamilov, Ulugbek S and Bouman, Charles A and Buzzard, Gregery T and Wohlberg, Brendt},
  journal={IEEE Signal Processing Magazine},
  volume={40},
  number={1},
  pages={85--97},
  year={2023},
  publisher={IEEE}
}

@article{maiden2009improved,
  title={An improved ptychographical phase retrieval algorithm for diffractive imaging},
  author={Maiden, Andrew M and Rodenburg, John M},
  journal={Ultramicroscopy},
  volume={109},
  number={10},
  pages={1256--1262},
  year={2009},
  publisher={Elsevier}
}

@article{maiden2012multislice,
  title={Ptychographic transmission microscopy in three dimensions using a multi-slice approach},
  author={Maiden, Andrew M and Humphry, Martin J and Rodenburg, John M},
  journal={Journal of the Optical Society of America a},
  volume={29},
  number={8},
  pages={1606--1614},
  year={2012},
  publisher={Optical Society of America}
}

@article{suzuki2014multislice,
  title={High-resolution multislice x-ray ptychography of extended thick objects},
  author={Suzuki, Akihiro and Furutaku, Shin and Shimomura, Kei and Yamauchi, Kazuto and Kohmura, Yoshiki and Ishikawa, Tetsuya and Takahashi, Yukio},
  journal={Physical review letters},
  volume={112},
  number={5},
  pages={053903},
  year={2014},
  publisher={APS}
}

@article{oleary2024buried,
  title={Three-dimensional structure of buried heterointerfaces revealed by multislice ptychography},
  author={O’Leary, Colum M and Sha, Haozhi and Zhang, Jianhua and Su, Cong and Kahn, Salman and Jiang, Huaidong and Zettl, Alex and Ciston, Jim and Miao, Jianwei},
  journal={Physical Review Applied},
  volume={22},
  number={1},
  pages={014016},
  year={2024},
  publisher={APS}
}

@article{perona1990scale,
  title={Scale-space and edge detection using anisotropic diffusion},
  author={Perona, Pietro and Malik, Jitendra},
  journal={IEEE Transactions on pattern analysis and machine intelligence},
  volume={12},
  number={7},
  pages={629--639},
  year={2002},
  publisher={IEEE}
}

@article{du2026fidelity,
  title={Fidelity-preserving enhancement of ptychography with foundational text-to-image models},
  author={Du, Ming and Rose, Volker and Deng, Junjing and Singh, Dileep and Chen, Si and Cherukara, Mathew},
  journal={Optica},
  volume={13},
  number={1},
  pages={45--56},
  year={2026},
  publisher={Optica Publishing Group}
}

@inproceedings{hoidn2023gridartifacts,
  title={Periodic artifacts generation and suppression in X-ray ptychography},
  author={Liu, Shilei and Xu, Zijian and Xing, Zhenjiang and Zhang, Xiangzhi and Li, Ruoru and Qin, Zeping and Wang, Yong and Tai, Renzhong},
  booktitle={Photonics},
  volume={10},
  number={5},
  pages={532},
  year={2023},
  organization={MDPI}
}

@incollection{huber1964robust,
  title={Robust estimation of a location parameter},
  author={Huber, Peter J},
  booktitle={Breakthroughs in statistics: Methodology and distribution},
  pages={492--518},
  year={1992},
  publisher={Springer}
}

@article{cao2022automatic,
  title={Automatic parameter selection for electron ptychography via Bayesian optimization},
  author={Cao, Michael C and Chen, Zhen and Jiang, Yi and Han, Yimo},
  journal={Scientific Reports},
  volume={12},
  number={1},
  pages={12284},
  year={2022},
  publisher={Nature Publishing Group UK London}
}

@article{umeike2025adapting,
  title={Adapting General-Purpose Foundation Models for X-ray Ptychography in Low-Data Regimes},
  author={Umeike, Robinson and Getty, Neil and Xiangyu, Yin and Jiang, Yi},
  journal={arXiv preprint arXiv:2511.02503},
  year={2025}
}

@ARTICLE{du2025ptychi,
  title         = "Pty-Chi: A {PyTorch}-based modern ptychographic data analysis
                   package",
  author        = "Du, Ming and Ruth, Hanna and Henke, Steven and Jiang, Yi and
                   Nikitin, Viktor and Tripathi, Ashish and Deng, Junjing and
                   Klug, Jeffrey and Myint, Peco and Zhou, Tao and Schwarz,
                   Nicholas and Cherukara, Mathew and Sandy, Alec and Vogt,
                   Stefan",
  journal       = "arXiv [physics.optics]",
  month         =  oct,
  year          =  2025,
  archivePrefix = "arXiv",
  primaryClass  = "physics.optics"
}

@ARTICLE{Rodenburg2004-nw,
  title   = "A phase retrieval algorithm for shifting illumination",
  author  = "Rodenburg, J M and Faulkner, H M L",
  journal = "Appl. Phys. Lett.",
  volume  =  85,
  number  =  20,
  pages   = "4795--4797",
  month   =  nov,
  year    =  2004
}

@ARTICLE{Maiden2009-md,
  title   = "An improved ptychographical phase retrieval algorithm for
             diffractive imaging",
  author  = "Maiden, Andrew M and Rodenburg, John M",
  journal = "Ultramicroscopy",
  volume  =  109,
  number  =  10,
  pages   = "1256--1262",
  year    =  2009
}

@ARTICLE{Maiden2017-um,
  title   = "Further improvements to the ptychographical iterative engine",
  author  = "Maiden, Andrew and Johnson, Daniel and Li, Peng",
  journal = "Optica",
  volume  =  4,
  number  =  7,
  pages   = "710--736",
  year    =  2017
}

@ARTICLE{Odstrcil2018-ns,
  title   = "Iterative least-squares solver for generalized maximum-likelihood
             ptychography",
  author  = "Odstr\v{c}il, Michal and Menzel, Andreas and Guizar-Sicairos,
             Manuel",
  journal = "Opt. Express",
  volume  =  26,
  number  =  3,
  pages   = "3108--3123",
  year    =  2018
}

@ARTICLE{Elser2003-ov,
  title   = "Phase retrieval by iterated projections",
  author  = "Elser, Veit",
  journal = "J. Opt. Soc. Am.",
  volume  =  20,
  number  =  1,
  pages   =  40,
  year    =  2003
}

@ARTICLE{Wakonig2020-ap,
  title   = "{PtychoShelves}, a versatile high-level framework for
             high-performance analysis of ptychographic data",
  author  = "Wakonig, Klaus and Stadler, Hans-Christian and Odstr\v{c}il, Michal
             and Tsai, Esther H R and Diaz, Ana and Holler, Mirko and Usov, Ivan
             and Raabe, J{\"{o}}rg and Menzel, Andreas and Guizar-Sicairos,
             Manuel",
  journal = "J. Appl. Crystallogr.",
  volume  =  53,
  number  =  2,
  year    =  2020
}

@ARTICLE{Nashed2014-eh,
  title   = "Parallel ptychographic reconstruction",
  author  = "Nashed, Youssef S G and Vine, David J and Peterka, Tom and Deng,
             Junjing and Ross, Rob and Jacobsen, Chris",
  journal = "Opt. Express",
  volume  =  22,
  number  =  26,
  pages   = "32082--32016",
  year    =  2014
}

@INPROCEEDINGS{Yue2021-ag,
  title     = "Ptychopy: {GPU} framework for ptychographic data analysis",
  author    = "Yue, Ke and Deng, Junjing and Jiang, Yi and Nashed, Youssef and
               Vine, David and Vogt, Stefan",
  editor    = "Lai, Barry and Somogyi, Andrea",
  booktitle = "X-Ray Nanoimaging: Instruments and Methods V",
  publisher = "SPIE",
  month     =  sep,
  year      =  2021
}

@ARTICLE{Enders2016-vs,
  title   = "A computational framework for ptychographic reconstructions",
  author  = "Enders, B and Thibault, P",
  journal = "Proceedings of the Royal Society A: Mathematical, Physical and
             Engineering Sciences",
  volume  =  472,
  number  =  2196,
  pages   =  20160640,
  year    =  2016
}

@ARTICLE{Favre-Nicolin2020-cj,
  title   = "{PyNX}: high-performance computing toolkit for coherent {X}-ray
             imaging based on operators",
  author  = "Favre-Nicolin, Vincent and Girard, Ga\'{e}tan and Leake, Steven and
             Carnis, Jerome and Chushkin, Yuriy and Kieffer, Jerome and Paleo,
             Pierre and Richard, Marie-Ingrid",
  journal = "J. Appl. Crystallogr.",
  volume  =  53,
  number  =  5,
  pages   = "1404--1413",
  year    =  2020
}

@ARTICLE{Paszke2019-xm,
  title   = "{PyTorch}: An Imperative Style, High-Performance Deep Learning
             Library",
  author  = "Paszke, Adam and Gross, Sam and Massa, Francisco and Lerer, Adam
             and Bradbury, James and Chanan, Gregory and Killeen, Trevor and
             Lin, Zeming and Gimelshein, Natalia and Antiga, Luca and Desmaison,
             Alban and K{\"{o}}pf, Andreas and Yang, Edward and DeVito, Zach and
             Raison, Martin and Tejani, Alykhan and Chilamkurthy, Sasank and
             Steiner, Benoit and Fang, Lu and Bai, Junjie and Chintala, Soumith",
  journal = "arXiv",
  year    =  2019
}

@ARTICLE{Huang2017-rz,
  title     = "Artifact mitigation of ptychography integrated with on-the-fly
               scanning probe microscopy",
  author    = "Huang, Xiaojing and Yan, Hanfei and Ge, Mingyuan and
               {\"{O}}zt{\"{u}}rk, Hande and Nazaretski, Evgeny and Robinson,
               Ian K and Chu, Yong S",
  journal   = "Appl. Phys. Lett.",
  publisher = "AIP Publishing",
  volume    =  111,
  number    =  2,
  pages     =  023103,
  month     =  jul,
  year      =  2017,
  language  = "en"
}

@article{aslan2021joint,
  title={Joint ptycho-tomography with deep generative priors},
  author={Aslan, Selin and Liu, Zhengchun and Nikitin, Viktor and Bicer, Tekin and Leyffer, Sven and G{\"u}rsoy, Do{\u{g}}a},
  journal={Machine Learning: Science and Technology},
  volume={2},
  number={4},
  pages={045017},
  year={2021},
  publisher={IOP Publishing}
}

@article{barutcu2022compressive,
  title={Compressive ptychography using deep image and generative priors},
  author={Barutcu, Semih and G{\"u}rsoy, Do{\u{g}}a and Katsaggelos, Aggelos K},
  journal={arXiv preprint arXiv:2205.02397},
  year={2022}
}

@article{seifert2023maximum,
  title={Maximum-likelihood estimation in ptychography in the presence of Poisson--Gaussian noise statistics},
  author={Seifert, Jacob and Shao, Yifeng and Van Dam, Rens and Bouchet, Dorian and Van Leeuwen, Tristan and Mosk, Allard P},
  journal={Optics Letters},
  volume={48},
  number={22},
  pages={6027--6030},
  year={2023},
  publisher={Optica Publishing Group}
}

@article{leidl2024influence,
  title={Influence of loss function and electron dose on ptychography of 2D materials using the Wirtinger flow},
  author={Leidl, Max Leo and Diederichs, Benedikt and Sachse, Carsten and M{\"u}ller-Caspary, Knut},
  journal={Micron},
  volume={185},
  pages={103688},
  year={2024},
  publisher={Elsevier}
}

@article{wu2024dose,
  title={Dose-efficient automatic differentiation for ptychographic reconstruction},
  author={Wu, Longlong and Yoo, Shinjae and Chu, Yong S and Huang, Xiaojing and Robinson, Ian K},
  journal={Optica},
  volume={11},
  number={6},
  pages={821--830},
  year={2024},
  publisher={Optica Publishing Group}
}

@inproceedings{kingma2015adam,
  author    = {Kingma, Diederik P. and Ba, Jimmy},
  title     = {Adam: A Method for Stochastic Optimization},
  booktitle = {International Conference on Learning Representations (ICLR)},
  year      = {2015},
  url       = {https://arxiv.org/abs/1412.6980}
}

@article{barzilai1988two,
  title={Two-point step size gradient methods},
  author={Barzilai, Jonathan and Borwein, Jonathan M},
  journal={IMA journal of numerical analysis},
  volume={8},
  number={1},
  pages={141--148},
  year={1988},
  publisher={Oxford University Press}
}

\appendix

\section{Best Discovered Regularization Algorithms}
\label{app:algorithms}

This appendix presents the complete source code of the best regularization algorithm discovered by \ptychi{} for each of the three datasets. Each algorithm is a drop-in \texttt{regularize\_llm(self)} method that operates on \texttt{self.data} (a complex-valued tensor of shape $[S, H, W]$ representing the object slices) and writes the regularized result back via \texttt{self.set\_data()}.

\subsection{Multislice X-ray (Algorithm 5efac846, SSIM 0.871)}
\label{app:multislice}

\begin{lstlisting}[basicstyle=\ttfamily\scriptsize,caption={Best discovered regularizer for the multislice X-ray dataset. Combines isotropic 3-D Charbonnier total variation, multi-slice gradient exclusion, Gram orthogonality, and complementary spectral masking, all updated via a stateful complex Adam optimizer with a three-phase learning rate schedule.}]
def regularize_llm(self):
    """Hybrid regulariser combining four gradient components
    with a stateful complex Adam optimizer."""
    import math, torch

    # ---- Shorthands and bookkeeping ----
    x = self.data                           # (S, H, W), complex
    dev, dt = x.device, x.dtype
    S, H, W = x.shape
    eps = 1e-8

    # Track the iteration count across calls
    it = getattr(self, "_iter_cnt", 0) + 1
    self._iter_cnt = it

    # Initialize Adam optimizer state on first call
    if not all(hasattr(self, k) for k in ("_adam_m", "_adam_v", "_adam_t")):
        self._adam_m = torch.zeros_like(x)        # first moment
        self._adam_v = torch.zeros_like(x.real)   # second moment (real)
        self._adam_t = 0
    self._adam_t += 1

    # Mixed-precision helper: use fp16 scratch buffers on CUDA
    def _mp(t):
        if (t.is_floating_point() and t.dtype == torch.float32
                and torch.cuda.is_available()):
            return t.to(torch.float16)
        return t

    # ---- Hyper-parameters ----
    LR_BASE   = 2.4e-2             # base learning rate
    LR_WARMUP = 5                  # linear warm-up iterations
    LR_DECAY  = 0.990              # exponential decay factor
    COS_START = 120                # cosine tail begins here
    b1, b2    = 0.9, 0.999         # Adam momentum parameters
    STEP_CLAMP = 0.08              # per-element step size clamp

    LAM_TV0   = 3.9e-3             # TV strength
    TV_K      = 1.2                # logistic steepness for TV weight
    LAM_DEPTH = 6.8e-3             # depth-axis TV weight
    D_CHAR    = 3.3e-2             # Charbonnier smoothing parameter
    LAM_EXCL  = 1.9e-3             # gradient exclusion strength
    LAM_CORR  = 2.9e-3             # Gram orthogonality strength
    LAM_SPEC0 = 1.4e-3             # spectral mask initial strength
    F_CUT     = 0.18               # Butterworth radial cutoff

    # ---- Finite differences (reflective boundary) ----
    dx = x - torch.cat([x[..., :, 1:], x[..., :, -1:]], dim=-1)
    dy = x - torch.cat([x[..., 1:, :], x[..., -1:, :]], dim=-2)
    if S > 1:
        dz = torch.zeros_like(x)
        dz[:-1] = x[:-1] - x[1:]    # reflective BC at last slice
    else:
        dz = torch.zeros_like(x)

    # ---- Adaptive contrast weight for TV ----
    # Blend reciprocal and logistic weighting maps,
    # clipped at the 5th and 95th percentiles
    g_xy = (dx.abs() + dy.abs()).mean(dim=(-2, -1), keepdim=True) + eps
    flat_c = g_xy.view(-1)
    p5, p95 = torch.quantile(
        flat_c, torch.tensor([0.05, 0.95], device=dev))
    g_clamp = g_xy.clamp(p5, p95)
    mu, sig = g_clamp.mean(), g_clamp.std().clamp(min=eps)
    w_rcp = (g_clamp.mean() / g_clamp).clamp(0.5, 2.0)
    w_log = torch.sigmoid(TV_K * (mu - g_clamp) / sig)
    tv_wgt = LAM_TV0 * 0.5 * (w_rcp + w_log)

    # ==== Component 1: Isotropic 3-D Charbonnier TV ====
    dz_mod = LAM_DEPTH * dz
    denom = torch.sqrt(
        dx.abs()**2 + dy.abs()**2 + dz_mod.abs()**2 + D_CHAR**2)
    w_iso = tv_wgt / denom
    tvx, tvy, tvz = _mp(w_iso * dx), _mp(w_iso * dy), _mp(w_iso * dz_mod)
    # Compute divergence (scatter-add)
    tv_grad = torch.zeros_like(x)
    tv_grad[..., :, :-1] += tvx[..., :, :-1]
    tv_grad[..., :,  1:] -= tvx[..., :, :-1]
    tv_grad[..., :-1, :] += tvy[..., :-1, :]
    tv_grad[...,  1:, :] -= tvy[..., :-1, :]
    if S > 1:
        tv_grad[:-1] += tvz[:-1]
        tv_grad[1:]  -= tvz[:-1]

    # ==== Component 2: Multi-slice gradient exclusion ====
    # Penalize co-located edges across slices (DDIP-style)
    excl_grad = torch.zeros_like(x)
    if S > 1 and LAM_EXCL > 0:
        g_mag = torch.sqrt(dx.real**2 + dy.real**2 + eps)
        share = g_mag.sum(dim=0, keepdim=True) - g_mag
        w_ex = _mp(LAM_EXCL * share / (g_mag + D_CHAR))
        dx_ex, dy_ex = w_ex * dx, w_ex * dy
        excl_grad[..., :, :-1] += dx_ex[..., :, :-1]
        excl_grad[..., :,  1:] -= dx_ex[..., :, :-1]
        excl_grad[..., :-1, :] += dy_ex[..., :-1, :]
        excl_grad[...,  1:, :] -= dy_ex[..., :-1, :]

    # ==== Component 3: Slice Gram-orthogonality ====
    # Minimize off-diagonal elements of the slice Gram matrix
    # to encourage independent slice content
    flat = x.view(S, -1)
    if S > 32:       # low-rank random projection for many slices
        k = min(32, max(8, S // 4))
        if not (hasattr(self, "_rand_proj")
                and self._rand_proj.shape == (k, S)):
            self._rand_proj = torch.randn(
                k, S, device=dev, dtype=flat.real.dtype)
        Y = self._rand_proj @ flat
        Gk = flat @ Y.T
        frob = torch.linalg.norm(Gk) + eps
        corr_grad = (2.0 * LAM_CORR / frob) * (Gk @ Y)
    else:            # full Gram matrix path
        G = flat.conj() @ flat.T
        G.fill_diagonal_(0.0)
        frob = torch.linalg.norm(G) + eps
        corr_grad = (2.0 * LAM_CORR / frob) * (G @ flat)
    corr_grad = corr_grad.view_as(x)

    # ==== Component 4: Complementary spectral masking ====
    # 6th-order Butterworth filter; low-contrast slices get
    # high-pass, others get low-pass. Strength is annealed.
    lam_spec = LAM_SPEC0 * math.exp(-max(0, it - 5) / 20.0)
    if lam_spec < 1e-6:
        spec_grad = torch.zeros_like(x)
    else:
        cache = getattr(self, "_freq_cache", None)
        if not (isinstance(cache, dict)
                and cache.get("shape") == (H, W)):
            fy = torch.fft.fftfreq(H, device=dev,
                                   dtype=x.real.dtype).view(-1, 1)
            fx = torch.fft.fftfreq(W, device=dev,
                                   dtype=x.real.dtype).view(1, -1)
            fr = torch.sqrt(fx**2 + fy**2) + eps
            low  = 1.0 / (1.0 + (fr / F_CUT)**6)
            high = 1.0 - low
            self._freq_cache = {
                "shape": (H, W), "low": low, "high": high}
        low_m = self._freq_cache["low"]
        high_m = self._freq_cache["high"]
        is_low = (g_clamp < g_clamp.mean()).to(x.real.dtype)
        mask = is_low * high_m + (1.0 - is_low) * low_m
        F = torch.fft.fft2(x)
        spec_grad = lam_spec * torch.fft.ifft2(mask * F)

    # ---- Aggregate gradient with robust clamping ----
    grad = tv_grad + excl_grad + corr_grad + spec_grad
    grad = torch.nan_to_num(grad)
    # Stage 1: scale-invariant median clamp
    med = torch.median(grad.abs()) + eps
    grad = grad * torch.clamp(
        4.0 * med / (grad.abs() + eps), max=1.0)
    # Stage 2: per-slice L2 norm clamp
    gn = torch.linalg.norm(
        grad.view(S, -1), dim=1, keepdim=True) + eps
    thr = 4.0 * torch.median(gn)
    grad = (grad.view(S, -1)
            * torch.clamp(thr / gn, max=1.0)).view_as(x)

    # ---- Complex Adam update ----
    m = self._adam_m = b1 * self._adam_m + (1 - b1) * grad
    v = self._adam_v = (b2 * self._adam_v
        + (1 - b2) * (grad.real**2 + grad.imag**2))
    m_hat = m / (1 - b1 ** self._adam_t)
    v_hat = v / (1 - b2 ** self._adam_t)
    # Three-phase LR: warm-up -> exponential decay -> cosine tail
    if it <= LR_WARMUP:
        lr = LR_BASE * it / LR_WARMUP
    else:
        lr = LR_BASE * (LR_DECAY ** (it - LR_WARMUP))
        if it > COS_START:
            cos_fac = 0.5 * (1 + math.cos(
                math.pi * (it - COS_START) / COS_START))
            lr *= max(cos_fac, 0.0)
    step = lr * m_hat / (torch.sqrt(v_hat) + eps)
    # Per-element step clamp for stability
    step = (torch.clamp(step.real, -STEP_CLAMP, STEP_CLAMP)
            + 1j * torch.clamp(step.imag, -STEP_CLAMP, STEP_CLAMP))

    x_new = x - step

    # ---- Per-slice L2 renormalization ----
    old_n = torch.linalg.norm(
        x.view(S, -1), dim=1, keepdim=True).clamp(min=eps)
    new_n = torch.linalg.norm(
        x_new.view(S, -1), dim=1, keepdim=True).clamp(min=eps)
    x_new = (x_new.view(S, -1) * (old_n / new_n)).view_as(x)
    x_new = torch.nan_to_num(x_new)

    self.set_data(x_new)
\end{lstlisting}

\subsection{X-ray Integrated Circuit (Algorithm 0165719f, SSIM 0.785)}
\label{app:ic}

\begin{lstlisting}[basicstyle=\ttfamily\scriptsize,caption={Best discovered regularizer for the X-ray integrated circuit dataset. Features an internal iterative optimization loop with separate amplitude (25 iterations) and phase (15 iterations) denoising sub-problems, structure-tensor-guided anisotropic Huber-TV, Barzilai--Borwein step size with Armijo acceptance, and automatic notch filtering of periodic artifacts.}]
def regularize_llm(self):
    """Multi-scale anisotropic DTGV-Huber regulariser
    with adaptive notch filtering."""
    import torch
    import torch.nn.functional as F

    # ---- Helper functions ----
    def _grad(u):
        """Forward finite differences along x and y."""
        gy, gx = torch.gradient(u, dim=(-2, -1))
        return gx, gy

    def _div(px, py):
        """Discrete divergence (adjoint of _grad)."""
        dpx_dx = torch.gradient(px, dim=(-2, -1))[1]
        dpy_dy = torch.gradient(py, dim=(-2, -1))[0]
        return dpx_dx + dpy_dy

    def _lap(u):
        """Discrete Laplacian."""
        return _div(*_grad(u))

    def _orientation_field(a, kernel):
        """Estimate local orientation from structure tensor."""
        gx, gy = _grad(a)
        conv = lambda t: F.conv2d(
            t.unsqueeze(1), kernel, padding=1).squeeze(1)
        J11, J22, J12 = conv(gx*gx), conv(gy*gy), conv(gx*gy)
        theta = 0.5 * torch.atan2(
            2.0 * J12, J11 - J22 + 1e-12)
        return torch.cos(theta), torch.sin(theta)

    def _build_notch_mask(a, k=8, sigma_frac=0.01):
        """Detect periodic artifacts from the power spectrum
        and build a multiplicative notch-rejection mask."""
        Hn, Wn = a.shape[-2], a.shape[-1]
        Famp = torch.fft.fftshift(
            torch.fft.fftn(a, dim=(-2, -1)), dim=(-2, -1))
        power = Famp.real**2 + Famp.imag**2
        # Zero out DC neighborhood
        power[Hn//2-1:Hn//2+2, Wn//2-1:Wn//2+2] = 0
        vals, idx = torch.topk(power.flatten(), k=k)
        sigma = sigma_frac * min(Hn, Wn)
        yy, xx = idx // Wn, idx % Wn
        yv, xv = torch.meshgrid(
            torch.arange(Hn, device=a.device),
            torch.arange(Wn, device=a.device), indexing='ij')
        H = torch.ones_like(power)
        thresh = power.mean() * 10
        for y0, x0, v in zip(yy, xx, vals):
            if v < thresh:
                continue
            d2 = (xv - x0)**2 + (yv - y0)**2
            g = torch.exp(-0.5 * d2 / sigma**2)
            H *= (1.0 - g)
            # Symmetric notch partner
            g2 = torch.roll(torch.roll(
                g, shifts=Hn//2, dims=0), shifts=Wn//2, dims=1)
            H *= (1.0 - g2)
        return H

    # ---- Setup ----
    z = self.data
    single = (z.ndim == 2)
    if single:
        z = z.unsqueeze(0)
    device = z.device
    c_dtype, r_dtype = z.dtype, torch.float32
    S, H, W = z.shape

    # Polar decomposition: separate amplitude and phase
    amp = torch.abs(z).to(r_dtype)
    phi = torch.angle(z).to(r_dtype)
    support = getattr(self, 'support_mask', None)
    if support is not None:
        if support.ndim == 2:
            support = support.unsqueeze(0)
        support = support.to(device=device, dtype=r_dtype)

    # ---- Hyper-parameters ----
    lam_tv   = 3e-3        # amplitude TV strength
    lam_tgv2 = 1e-3        # second-order TGV strength
    lam_phi  = 8e-4        # phase TV strength
    lam_cplx = 5e-4        # complex TV strength
    lam_fft  = 2e-3        # notch filter strength
    n_amp_iter  = 25       # amplitude sub-iterations
    n_phi_iter  = 15       # phase sub-iterations
    n_cplx_iter = 2        # complex TV iterations
    tau_amp0 = 0.24        # initial amplitude step size
    tau_phi0 = 0.24        # initial phase step size
    huber_eps0 = 1e-3      # Huber threshold (annealed)
    anisotropy = 4.0       # directional anisotropy ratio
    orient_every = 5       # re-estimate orientation interval
    tol = 5e-4             # early stopping tolerance
    device_eps = torch.tensor(1e-12, dtype=r_dtype, device=device)

    # Gaussian smoothing kernel for structure tensor
    k = torch.tensor(
        [[1,2,1],[2,4,2],[1,2,1]],
        dtype=r_dtype, device=device).view(1,1,3,3) / 16.0

    # Build notch mask from amplitude power spectrum
    notch_H = _build_notch_mask(amp[0], k=8).to(r_dtype)

    # ==== Amplitude denoising (25 iterations) ====
    # Anisotropic Huber-TV guided by structure tensor,
    # with Barzilai-Borwein step size and Armijo acceptance
    huber_eps = torch.tensor(huber_eps0, dtype=r_dtype, device=device)
    tau_amp = torch.tensor(tau_amp0, dtype=r_dtype, device=device)
    amp_prev = upd_prev = None
    vx, vy = _orientation_field(amp, k)

    for it in range(n_amp_iter):
        # Re-estimate local orientation periodically
        if it % orient_every == 0:
            vx, vy = _orientation_field(amp, k)

        gx, gy = _grad(amp)
        g_norm = torch.sqrt(gx*gx + gy*gy + device_eps)

        # Decompose gradient into parallel and perpendicular
        # components relative to the local edge direction
        g_par  = vx*gx + vy*gy
        g_perp = -vy*gx + vx*gy
        alpha = torch.abs(g_perp) / (
            torch.abs(g_par) + torch.abs(g_perp) + device_eps)
        coeff = 1.0 + (anisotropy - 1.0) * alpha

        # Huber-weighted anisotropic TV gradient
        px = gx / (coeff * g_norm + huber_eps)
        py = gy / (coeff * g_norm + huber_eps)
        upd = lam_tv * _div(px, py) - lam_tgv2 * _lap(amp)
        amp_new = amp + tau_amp * upd

        # Barzilai-Borwein step size with Armijo acceptance
        if amp_prev is not None:
            s = (amp_new - amp_prev).flatten()
            y = (upd - upd_prev).flatten()
            sy = torch.dot(s, y)
            yy = torch.dot(y, y) + device_eps
            tau_bb = torch.clamp(sy / yy, 1e-4, 0.25)
            # Accept BB step only if it reduces energy
            E_old = (lam_tv * g_norm.mean()
                     + lam_tgv2 * _lap(amp).abs().mean())
            amp_try = amp + tau_bb * upd
            gx_t, gy_t = _grad(amp_try)
            E_new = (lam_tv * torch.sqrt(
                         gx_t**2 + gy_t**2 + device_eps).mean()
                     + lam_tgv2 * _lap(amp_try).abs().mean())
            if E_new < E_old:
                tau_amp = tau_bb

        amp_prev, upd_prev = amp, upd
        amp = amp_new
        if support is not None:
            amp = amp * support
        # Anneal Huber threshold every 4 iterations
        if (it + 1) % 4 == 0:
            huber_eps *= 0.9
        # Early stopping on relative update magnitude
        rel = (torch.mean(torch.abs(upd))
               / (torch.mean(torch.abs(amp)) + device_eps))
        if rel < tol:
            break

    amp = torch.clamp(amp, min=0.0)

    # ==== Phase denoising (15 iterations) ====
    # Amplitude-weighted Huber-TV on the phase channel
    tau_phi = tau_phi0
    huber_eps = huber_eps0
    w_phi = lam_phi / (amp + 1e-6)  # weight inversely by amplitude

    for it in range(n_phi_iter):
        phx, phy = _grad(phi)
        ph_norm = torch.sqrt(phx**2 + phy**2 + device_eps)
        qx = w_phi * phx / (ph_norm + huber_eps)
        qy = w_phi * phy / (ph_norm + huber_eps)
        phi_new = phi + tau_phi * _div(qx, qy)
        # Wrap phase to [-pi, pi]
        phi = (phi_new + torch.pi) % (2.0*torch.pi) - torch.pi
        if (it + 1) % 4 == 0:
            huber_eps *= 0.9
        rel = (torch.mean(torch.abs(phi - phi_new))
               / (torch.mean(torch.abs(phi)) + device_eps))
        if rel < tol:
            break

    # ==== Joint complex TV (2 iterations) ====
    z = torch.polar(amp, phi).to(c_dtype)
    tau_c = 0.24
    for _ in range(n_cplx_iter):
        r, i = z.real.to(r_dtype), z.imag.to(r_dtype)
        rx, ry = _grad(r)
        ix, iy = _grad(i)
        rn = torch.sqrt(rx**2 + ry**2 + device_eps)
        in_ = torch.sqrt(ix**2 + iy**2 + device_eps)
        r = r + tau_c * lam_cplx * _div(
            rx/(rn+device_eps), ry/(rn+device_eps))
        i = i + tau_c * lam_cplx * _div(
            ix/(in_+device_eps), iy/(in_+device_eps))
        z = r.to(c_dtype) + 1j * i.to(c_dtype)

    # ==== Notch filtering of periodic artifacts ====
    if lam_fft > 0:
        Zf = torch.fft.fftn(z, dim=(-2, -1))
        Zf_shift = torch.fft.fftshift(Zf, dim=(-2, -1))
        Zf_shift *= (1.0 - lam_fft * notch_H).to(c_dtype)
        z = torch.fft.ifftn(
            torch.fft.ifftshift(Zf_shift, dim=(-2, -1)),
            dim=(-2, -1))

    # Softplus amplitude activation and support masking
    beta = 1.0 / (amp.std(unbiased=False) + 1e-6)
    z_amp = F.softplus(torch.abs(z), beta=beta)
    z = torch.polar(z_amp.to(r_dtype), torch.angle(z)).to(c_dtype)
    if support is not None:
        z = z * support.to(c_dtype)

    if single:
        self.set_data(z[0])
    else:
        self.set_data(z)
\end{lstlisting}

\subsection{Apoferritin (Algorithm 6d2c9f3f, SSIM 0.836)}
\label{app:apoferritin}

\begin{lstlisting}[basicstyle=\ttfamily\scriptsize,caption={Best discovered regularizer for the apoferritin dataset. Implements a sequential signal-processing pipeline: noise-adaptive soft-Huber shrinkage, cascaded guided filtering, Itoh phase unwrapping, Perona--Malik anisotropic diffusion, and a weak L2 anchor to the original estimate.}]
def regularize_llm(self):
    """Hybrid regulariser -- crossover of two parental variants.
    Sequential signal-processing pipeline operating in polar
    (amplitude/phase) coordinates."""
    import math
    import torch
    import torch.nn.functional as F

    data0 = self.data                       # [S, H, W], complex
    dev, cdt = data0.device, data0.dtype
    rdt = data0.real.dtype
    eps = 1e-6
    two_pi = 2.0 * math.pi
    S, H, W = data0.shape

    # ---- Utility functions ----
    def _box(x, r):
        """Fast 2-D box mean via average pooling."""
        if r <= 0:
            return x
        return F.avg_pool2d(x, 2*r+1, stride=1, padding=r)

    def _grad(u):
        """Forward differences via circular shift."""
        gx = torch.roll(u, -1, 2) - u
        gy = torch.roll(u, -1, 1) - u
        return torch.stack((gx, gy), dim=0)

    def _div(p):
        """Discrete divergence (adjoint of _grad)."""
        px, py = p[0], p[1]
        return ((px - torch.roll(px, 1, 2))
                + (py - torch.roll(py, 1, 1)))

    def _Lc(u, c):
        """Anisotropic diffusion operator: div(c * grad(u))."""
        return _div(c * _grad(u))

    # ==== Step 1: Polar decomposition ====
    amp = data0.abs()
    phi = torch.angle(data0)

    # ==== Step 2: Noise-adaptive soft-Huber shrinkage ====
    # Shrink amplitude toward 1.0 with data-driven threshold
    glob_var = amp.var(unbiased=False)
    k_min, k_max = 0.07, 0.20
    k_soft = k_min + (k_max - k_min) * torch.sigmoid(
        (glob_var - 0.02) * 25.0)
    delta = amp - 1.0
    amp = 1.0 + delta * (k_soft / (delta.abs() + k_soft + eps))

    # ==== Step 3: Cascaded guided filtering ====
    # Multi-radius edge-preserving smoothing driven by
    # local variance
    for r in (1, 2):
        mean_r = _box(amp.unsqueeze(1), r).squeeze(1)
        var_r = _box(
            (amp - mean_r).pow(2).unsqueeze(1), r).squeeze(1)
        w_r = 1.0 / (var_r + eps)
        amp = (mean_r + w_r * amp) / (1.0 + w_r)

    # Optional 3x3 Gaussian smoothing for high-noise regime
    if glob_var > 0.1:
        gk = torch.tensor(
            [[1,2,1],[2,4,2],[1,2,1]],
            dtype=rdt, device=dev).reshape(1,1,3,3) / 16.0
        amp = F.conv2d(
            amp.unsqueeze(1), gk, padding=1).squeeze(1)

    # ==== Step 4: 2-D Itoh phase unwrapping ====
    # Sequential row-then-column pass per slice
    phi_u = phi.clone()
    for s in range(S):
        # Unwrap along rows
        d_row = phi_u[s, :, 1:] - phi_u[s, :, :-1]
        d_row = (d_row + math.pi) % two_pi - math.pi
        phi_u[s, :, 1:] = phi_u[s, :, :-1] + d_row
        # Unwrap along columns
        d_col = phi_u[s, 1:, :] - phi_u[s, :-1, :]
        d_col = (d_col + math.pi) % two_pi - math.pi
        phi_u[s, 1:, :] = phi_u[s, :-1, :] + d_col

    # Remove DC/ramp by subtracting slice mean
    phi_u = phi_u - phi_u.mean(dim=(1, 2), keepdim=True)

    # ==== Step 5: Perona-Malik anisotropic diffusion ====
    # Edge-preserving smoothing of the unwrapped phase
    lam_pm = 0.1       # contrast parameter
    for _ in range(5):
        grad_u = _grad(phi_u)
        mag2 = grad_u[0].pow(2) + grad_u[1].pow(2)
        # Conduction coefficient: preserves edges
        c = 1.0 / (1.0 + mag2 / (lam_pm**2) + eps)
        phi_u = phi_u + 0.1 * _Lc(phi_u, c)

    # ==== Step 6: L2 anchor to original estimate ====
    # Blend regularized result with original to prevent
    # over-smoothing (90% regularized, 10% original)
    alpha = 0.1
    data_new = torch.polar(amp, phi_u)
    data_out = alpha * data0 + (1.0 - alpha) * data_new.to(cdt)

    self.set_data(data_out)
\end{lstlisting}

\end{document}